\documentclass[british,english]{article}
\usepackage[T1]{fontenc}
\usepackage[latin9]{inputenc}
\usepackage{geometry}
\usepackage{amsmath}
\usepackage{amssymb}
\usepackage{graphicx}
\usepackage{caption}
\usepackage{subcaption}
\usepackage[authoryear]{natbib}

\usepackage{float}

\usepackage{babel}

\usepackage{algpseudocode}
\usepackage{algorithm}

\begin{document}

\title{Delayed acceptance ABC-SMC}

\author{Richard G. Everitt and Paulina A. Rowi{\'n}ska}
\maketitle
\begin{abstract}
Approximate Bayesian computation (ABC) is now an established technique
for statistical inference used in cases where the likelihood function
is computationally expensive or not available. It relies on the use
of a~model that is specified in the form of a~simulator, and approximates
the likelihood at a~parameter value $\theta$ by simulating auxiliary data
sets $x$ and evaluating the distance of $x$ from the true data $y$.
However, ABC is not computationally feasible in cases where using
the simulator for each $\theta$ is very expensive. This paper investigates
this situation in cases where a~cheap, but approximate, simulator
is available. The approach is to employ delayed acceptance Markov
chain Monte Carlo (MCMC) within an ABC sequential Monte Carlo (SMC)
sampler in order to, in a~first stage of the kernel, use the cheap
simulator to rule out parts of the parameter space that are not worth
exploring, so that the ``true'' simulator is only run (in the second
stage of the kernel) where there is a~reasonable chance of accepting proposed
values of $\theta$. We show that this approach can be used quite
automatically, with few tuning parameters. Applications to stochastic differential
equation models and latent doubly intractable distributions are presented.
\end{abstract}

\section{Introduction}

\subsection{Motivation}

Approximate Bayesian computation (ABC) is a~technique for approximate
Bayesian inference originally introduced in the population genetics
literature \citep{Pritchard1999,Beaumont2002}, but which is now used
for a~wide range of applications. It is suitable for situations
in which a~model $l\left(\cdot\mid\theta\right)$ with parameters
$\theta$ for data $y$ is readily available in the form of a~simulator,
but where $l$ as a~function of $\theta$ (known as the likelihood)
is intractable in that it cannot be evaluated pointwise. If $\theta$
is assigned prior distribution $p$, the object of inference is the
posterior distribution $\pi\left(\theta\mid y\right)\propto p\left(\theta\right)l\left(y\mid\theta\right)$.
ABC yields approximations to the posterior distribution through approximating
the likelihood, in the simplest case using

\begin{equation}
\widehat{f_{\epsilon}}\left(y\mid\theta\right)=\frac{1}{M}\sum_{m=1}^{M}P_{\epsilon}\left(y\mid x^{(m)}\right).\label{eq:abc_approx_mc}
\end{equation}
where $P_{\epsilon}$ is a~kernel centred around $x^{(m)}$ and $M$ points
$\left\{ x^{(m)}\right\} _{m=1}^{M}$  are sampled from $l\left(\cdot\mid\theta\right)$. We may see Monte Carlo ABC algorithms as sampling from a~joint distribution
on $\theta$ and $x$ proportional to $p\left(\theta\right)l\left(x\mid\theta\right)P_{\epsilon}\left(y\mid x\right)$,
where $l$
is used as a~proposal distribution for $x$.

In this paper we describe methods that are designed to be applicable
in cases where the likelihood estimation is computationally costly
because the model $l$ is expensive to simulate from, for example
when studying an ordinary differential equation (ODE) or stochastic
differential equation (SDE) model \citep{Picchini2014} that requires
a solver with a~small step size, or when using an individual based
model \citep{VanderVaart2015} with a~large number of individuals.
This situation is not uncommon, since even if the model takes only
a matter of seconds to simulate, this cost is prohibitive when it
needs to be simulated a~number of times due to estimating the likelihood
at a~large number of Monte Carlo points in $\theta$-space. We note
that this situation is exacerbated when $\theta$ is of moderate to
high dimension, since in these cases (as in any Monte Carlo method)
it is difficult to design a~proposal that can efficiently make moves
on all dimensions of $\theta$ simultaneously. Most commonly, each
dimension of $\theta$ is updated using a~single component move, and
in the ABC context this means using $M$ simulations from the likelihood
when updating each dimension of $\theta$.

\subsection{Previous work and contributions}

This paper describes an ABC-SMC algorithm that uses a delayed-acceptance ABC-MCMC move that aims to limit the number of $\theta$ points at which we need to perform an expensive simulation of the likelihood, whilst also attempting to maintain a good exploration of the target distribution. We begin in section \ref{subsec:ABC-SMC} with a review of the literature on ABC-SMC, focussing particularly on the case where a uniform ABC kernel is used. Then in section \ref{subsec:Delayed-acceptance-with} we describe delayed-acceptance
ABC-MCMC for exploring $\theta$-space, and use it as a~method for discarding $\theta$ points that are unlikely to
be in regions of high posterior density. Delayed acceptance (DA) decomposes
an MCMC move into two stages. At the first stage a~point is proposed
and is accepted or rejected according to a~posterior that is usually
chosen to approximate the desired posterior. If accepted at the first
stage, at the second stage an acceptance probability is used that
``corrects'' for the discrepancy between the approximate and the
desired target. Thus we may think of the first stage as screening
for points to take through to the second stage. Compared to the previous
work, this approach is most similar to the ``early rejection'' of \citet{Picchini2013} who use the prior for this screening step.
Delayed-acceptance ABC-MCMC at the first stage makes use of an
approximate, but computationally cheap, alternative to the full simulator.
We will see that our approach is applicable in cases in which there
is a~cheap, approximate simulator is used independently of the
full simulator, and also where the full simulator
is a~continuation of the initial cheap simulation. This latter situation is the same as that considered in ``Lazy'' ABC \citep{Prangle2014}, in which is it possible to monitor a~simulation
as it is progressing, and to be able to assess before the full simulation
whether the current $\theta$ point is likely to be in a~region of
high posterior density.
We will see that using delayed-acceptance ABC-MCMC introduces an additional
tuning parameter: the tolerance $\epsilon$ used for the cheap simulator.
When using delayed-acceptance ABC-MCMC directly, it is not always
straightforward to set this parameter. In section \ref{subsec:Use-in-ABC-SMC}
we embed delayed-acceptance ABC-MCMC in the ABC-SMC framework, since
this is established as an effective method for guiding Monte
Carlo points in $\theta$-space to regions of high posterior density,
and also since we see that it allows for automating the choice of
the additional tuning parameter. The resultant algorithm is particularly useful for distributions for which
it is difficult to design a~useful proposal distribution. When our
proposal has a~poor acceptance rate, we may need to run the expensive
simulator many times for a~single acceptance. However, with delayed
acceptance we may propose a~very large number of points without the
need to run the expensive simulator for a large number of them. The procedure is to evaluate whether
the proposed points are in a~high density region according to the cheap
simulator -- in the cases where they are, we run the expensive simulator. For points that lie in this region (i.e. that pass the
first stage of DA), we must run the expensive simulator, and
expect to accept a~reasonable proportion of these (as long as the
cheap simulator is close to the expensive simulator). Essentially,
the first stage of the DA acts to give us a~well-targeted
proposal.
Section \ref{sec:Application-to-stochastic} considers an application
to SDEs, where we examine using cheap simulators with a~large
step size in the numerical solver. Section \ref{sec:Application-to-Ising}
considers applications to doubly intractable distributions: applying
ABC to these models (and other Markov random fields) is complicated because no computationally feasible exact simulator is available.
Instead an MCMC algorithm is commonly used as an approximate approach
\citep{Grelaud2009,Everitt2012}. In such approaches the burn in of
this MCMC method determines the accuracy of the approach: for a~small
burn in the method is biased, with this bias being eliminated by a
large burn in which may be computationally expensive. In the main paper we examine the case of latent Ising models previously studied in \citet{Everitt2012}, noting the potential
advantage of ABC in these cases compared to methods such as the exchange
algorithm. In the appendix, we also examine an application to a latent exponential random graph model (ERGM). We then conclude with a~discussion in section \ref{sec:Conclusions}.

\section{Background on ABC-SMC} \label{subsec:ABC-SMC}

\subsection{SMC samplers}

An SMC sampler \citep{DelMoral2006c} is an importance sampling based
Monte Carlo method for generating weighted points from~a target distribution
$\pi$ through making use of~a sequence of $T$ target distributions
that bridge between some initial proposal distribution $\pi_{0}$
from which we can simulate, and the final target $\pi_{T}=\pi$. Here
we focus on the variant of the algorithm that uses an MCMC kernel,
and provide~a sketch of the main steps of the algorithm (further details
may be found in \citet{DelMoral2006c}). The algorithm begins by simulating
$N$ weighted ``particles'' from $\pi_{0}$ (initially with equal
weights). Then the particles are moved from target $\pi_{t}$ to $\pi_{t+1}$
through an importance sampling reweighting step, followed by~a ``move''
step in which each particle is simulated from an MCMC kernel starting
at its current value, and with target $\pi_{t+1}$. Let $\theta_{t}^{(i)}$
be the value taken by the $i$-th particle after iteration $t$ and
$w_{t}^{(i)}$ be its (normalised) weight. The reweighting step then
computes the (unnormalised) weight $\tilde{w}_{t+1}^{(i)}$ of the
$n$-th particle $\theta_{t}^{(i)}$ using
\begin{eqnarray}
\tilde{w}_{t+1}^{(i)} & = & w_{t}^{(i)}\frac{\pi_{t+1}\left(\theta_{t}^{(i)}\right)}{\pi_{t}\left(\theta_{t}^{(i)}\right)},\label{eq:smc_weight-1}
\end{eqnarray}
followed by~a normalisation step to find $w_{t+1}^{(i)}$.
This algorithm yields an empirical approximation of $\pi_{t}$
\begin{equation}
\hat{\pi}_{t}^{N}=\sum_{i=1}^{N}w_{t}^{(i)}\delta_{\theta_{t}^{(i)}},\label{eq:smc_empirical_approx}
\end{equation}
where $\delta_{\theta}$ is~a Dirac mass at $\theta$, and an estimate
of its normalising constant
\begin{equation}
\hat{Z_{t}}=\prod_{s=1}^{t}\sum_{i=1}^{N}w_{s}^{(i)}\frac{\pi_{s+1}\left(\theta_{s}^{(i)}\right)}{\pi_{s}\left(\theta_{s}^{(i)}\right)}.\label{eq:smc_Z_est}
\end{equation}
Usually the variance of the weights is monitored
during the algorithm, and if this becomes too large (which would lead
to high variance estimators if no intervention was made)~a resampling
step is performed after reweighting (in this paper stratified resampling is used). The resampling step at iteration
$t+1$ simulates $N$ particles from the empirical distribution $\hat{\pi}_{t+1}^{N}$.

SMC outperforms importance sampling in cases where one needs many
points to obtain low variance importance sampling estimates when directly
using $\pi_{0}$ as~a proposal for $\pi$, i.e. when the distance
between $\pi_{0}$ and $\pi$ is too large. In order to ensure that SMC estimates have low variance, we
require the distance between $\pi_{t}$ and $\pi_{t+1}$ to be small
for all $t$.~a proxy for the distance between subsequent targets
that may be estimated as the algorithm runs is given by the \emph{effective
sample size} (ESS) \citep{Kong1994}
\begin{equation}
\mbox{ESS}=\left(\sum_{i=1}^{N}\left(w_{t+1}^{(i)}\right)^{2}\right)^{-1}.\label{eq:ess}
\end{equation}
A large ESS corresponds to~a small distance between targets although,
as referred to in the following section,~a high ESS is necessary but
not sufficient for achieving~a low Monte Carlo variance.

\subsection{ABC-SMC\label{subsec:Use-in-ABC}}

\citet{Sisson2007d,Sisson2009} remark that the ABC posterior $\pi_{\epsilon}$
is~a natural candidate for simulating from using SMC, in that~a sequence
of posteriors with~a decreasing sequence of tolerances $\epsilon_{1}$
to $\epsilon_{T}=\epsilon$ is~a natural and useful choice as~a sequence
of distributions to use in SMC. We follow \citet{DelMoral2012g} in
choosing the sequence of distributions
\begin{equation}
\pi_{t}\left(\theta_{t},x_{t}\mid y\right) \propto p\left(\theta_{t}\right)l\left(x_{t}\mid\theta_{t}\right)P_{\epsilon_{t}}\left(y\mid x_{t}\right)\label{eq:abc_smc_seq}
\end{equation}
such that $\theta_{t}\sim\pi_{\epsilon_{t}}$, with $x_{t}$ being
the corresponding auxiliary variable (as in equation (\ref{eq:abc_approx_mc}))
at iteration $t$. When an ABC-MCMC move is used for the ``move'' step, this leads to the following weight update when
moving from target $\pi_{t}$ to $\pi_{t+1}$
\begin{equation}
\tilde{w}_{t+1}^{(i)}=w_{t}^{(i)}\frac{P_{\epsilon_{t+1}}\left(y\mid x_{t}^{(i)}\right)}{P_{\epsilon_{t}}\left(y\mid x_{t}^{(i)}\right)}\label{eq:reweight_abc_smc}
\end{equation}
(a derivation may be found in the appendix). This weight update is computationally cheap and requires no simulation
to produce $x_{t}^{(i)}$, which has been generated either in the
initial step of the algorithm or as a~part of~a previous MCMC move.
Additionally, it is cheap to compute for any choice of $\epsilon_{t+1}$. This fact is used by \citet{DelMoral2012g} in order to devise
an adaptive ABC-SMC algorithm that automatically determines the sequence
$\left(\epsilon_{t}\right)$. At every SMC iteration~a bisection method
is used to determine the $\epsilon_{t+1}$ that will result in an
ESS that is some proportion (e.g. 90\%) of $N$.

It is common practice in SMC algorithms to use the current set of particles to adaptively
set the proposal for the MCMC kernels used in the ``move''
step (note that this introduces a small bias into estimates based on the SMC for the methods used in this paper). The simplest scheme, from \citet{Robert2011b}, sets the proposal covariance to
be some multiple of the empirical covariance of the particles (this
choice being rooted in results on optimal proposals for some MCMC
algorithms).

\subsubsection{ABC-SMC with indicator potentials\label{subsec:Alive-SMC}}

In ABC~a common choice for the kernel $P_{\epsilon_{t}}$ is $P_{\epsilon_{t}}\left(y\mid x_{t}\right)\propto\mathbb{I}\left(d\left(y,x_{t}\right)<\epsilon_{t}\right)$,
where $d$ is~a distance metric. This kernel is known to be sub-optimal \citep{Li2018a}, but we restrict our attention to this case for the remainder of the paper since it simplifies the presentation and interpretation of the algorithms we introduce. The first example of this simplification is that when used in the ABC-SMC method of \citet{DelMoral2012g}, the particle weights are either zero (``dead'') or non-zero (``alive'') (this following from equation (\ref{eq:reweight_abc_smc})). This type of SMC algorithm is also studied in \citet{Cerou2012,DelMoral2015,Prangle2016}. A further consequence that the ESS is equal to the number of ``alive''
particles with non-zero weight after the update, simplifying the interpretation of the adaptive approach to choosing $\epsilon_{t+1}$ described above.

In this paper we use the revised scheme of \citet{Bernton2017}. The revision addresses the issue
that the acceptance rate of ABC-MCMC moves decreases as the ABC-SMC
progresses (since the ABC tolerance decreases). When the MCMC moves
have~a poor acceptance rate, the ESS is not~a good criterion for deciding
on~a new tolerance, since it is unaffected by the values taken by
each particle: even if all particles take the same value, the ESS
may be high. Therefore \citet{Bernton2017} suggest instead to choose
$\epsilon_{t+1}$ such that some number $U$ (with $0<U\leq N$) of the
$N$ particles will be unique after resampling has been performed. Both this new scheme, and the original method that uses the ESS, introduce a small bias into estimates from the SMC \citep{Cerou2012,Prangle2016}.

To simplify the presentation of our method, we perform resampling at every step of the algorithm.  As~a consequence, the denominator in
equation (\ref{eq:reweight_abc_smc}) is positive and the same for every
particle, thus in practice the reweighting step becomes
\[
\tilde{w}_{t+1}^{(i)}=w_{t}^{(i)}\mathbb{I}\left(d\left(y,x_{t}^{(i)}\right)<\epsilon_{t+1}\right),
\]
where for simplicity we have omitted the additional normalising term for
$\mathbb{I}\left(d\left(y,x_{t}\right)<\epsilon_{t}\right)$, required in order to provide~a correct marginal likelihood estimator \---\ see \citet{Didelot2011} for details. Resampling at every step avoids propagating ``dead'' particles with zero weight, but is not an optimal strategy since it is possible that not all alive particles will be part of the resample (unless systematic resampling is used). A standard approach to deciding when to resample is to only do so when the ESS fall below a pre-specified threshold \citep{DelMoral2012}, but we do not consider such approaches in this paper.

The ABC-SMC algorithm with the configuration described in this section is given in algorithm \ref{alg:abc_smc}. Note that although a final tolerance $\epsilon_{\mbox{end}}$ is specified in this algorithm, in practice the method is run for as long as computational resources will allow; a standard approach would be to monitor the acceptance probability of the MCMC moves and to terminate when this is zero for a number of iterations. The early rejection method of \citet{Picchini2013} may be employed for the ABC-MCMC moves when using an indicator kernel: this approach is precisely the same as standard ABC-MCMC, with the calculation reorganised in order to avoid unnecessary simulations from the likelihood (see appendix for details).

\begin{algorithm}[H]
\caption{ABC-SMC using early rejection.}
\label{alg:abc_smc}
\hspace*{\algorithmicindent} \textbf{Inputs:} Number of particles $N$, desired number of unique particles $U$, prior $p$, simulator $l$, final tolerance $\epsilon_{\mbox{end}}$. \\
 \hspace*{\algorithmicindent} \textbf{Outputs:} Particles $\left\{\left( \theta^{(i)}_t,x^{(i)}_t \right) \right\}_{i=1}^N$ and weights $\left\{w_{t}^{(i)} \right\}_{i=1}^N$ for all $t$.
\begin{algorithmic}
\For {$i= 1:N$}
    \State $\theta^{(i)}_0 \sim p\left( \cdot \right)$
    \State $x^{(i)}_{0} \sim l\left( \cdot \mid \theta^{(i)}_0 \right)$
    \State $w^{(i)}_0 = 1/N$
\EndFor
\State $\epsilon_0 = \max_i d\left(y,x_{0}^{(i)}\right)$, $t=0$.
\While {$\epsilon_t > \epsilon_{\mbox{end}}$}
    \State Simulate $v \sim \mathcal{U}\left[ 0,1 \right]^N$, to be used in resampling.
    \State Use bisection to choose $\epsilon_{t+1}$ s.t. there will be $U$ unique particles after reweighting and resampling (using random numbers $v$).
    \For {$i= 1:N$}
        \State $\tilde{w}_{t+1}^{(i)}=w_{t}^{(i)}\mathbb{I}\left(d\left(y,x_{t}^{(i)}\right)<\epsilon_{t+1}\right)$
    \EndFor
    \State Normalise $\left\{ \tilde{w}_{t+1} \right\}_{i=1}^N$ to give normalised weights $\left\{ w_{t+1} \right\}_{i=1}^N$.
    \State Perform resampling using random draws $v$.
    \For {$i= 1:N$}
        \State $\theta^{(i)}_{t+1} =  \theta^{(i)}_{t}$, $x^{(i)}_{t+1} =  x^{(i)}_{t}$
        \State $\left(\theta^{(i)}_{t+1}\right)^* \sim q\left( \cdot \mid \theta^{(i)}_{t} \right)$
        \State $u \sim \mathcal{U}\left(0,1\right)$
        \If {$u < \frac{p\left( \left(\theta^{(i)}_{t+1}\right)^* \right) q\left( \theta^{(i)}_{t} \mid \left(\theta^{(i)}_{t+1}\right)^* \right) }{ p\left( \theta^{(i)}_{t+1} \right) q\left( \left(\theta^{(i)}_{t+1}\right)^* \mid \theta^{(i)}_{t} \right) }$ }
            \State $ \left(x^{(i)}_{t+1} \right)^* \sim l\left( \cdot \mid \left(\theta^{(i)}_{t+1}\right)^* \right)$
            \If { $d\left(y, \left(x^{(i)}_{t+1} \right)^* \right) < \epsilon_{t+1}$ }
                \State $\theta^{(i)}_{t+1} =  \left(\theta^{(i)}_{t+1}\right)^*$, $x^{(i)}_{t+1} =  \left(x^{(i)}_{t+1}\right)^*$
            \EndIf
        \EndIf
    \EndFor
    \State $t = t + 1$
\EndWhile
\end{algorithmic}
\end{algorithm}

\section{Delayed acceptance ABC-SMC\label{sec:Delayed-acceptance-ABC-SMC}}

This section describes the algorithm that is introduced by this paper: an ABC-SMC sampler that uses a DA ABC-MCMC kernel as its ``move'' step (instead of standard ABC-MCMC), with the DA move being tuned automatically using the population of particles. We begin by describing delayed acceptance, then outline how it may be used in the ABC context, before describing an ABC-SMC sampler that makes use of it. Where they are omitted from the main paper, full derivations are given in the appendix.

\subsection{Delayed acceptance\label{subsec:Delayed-acceptance}}

This section introduces delayed acceptance \citep{Christen2005} as the algorithm that results when one Metropolis-Hastings kernel
is used as a~proposal within another. We note here the link with pseudo-marginal
type algorithms that, as a~special case, show how to use importance
sampling or SMC within a~Metropolis-Hastings algorithm.
Let Metropolis-Hastings transition kernel $K_{1}$ have invariant
distribution $\pi_{1}$ and suppose our aim is to use the distribution
$\pi_{1}$ as the proposal distribution in another Metropolis-Hastings
kernel $K_{2}$ with an~invariant distribution $\pi_{2}$. It turns out
that we can use a~draw from $K_{1}$ as if it was a
point drawn from $\pi_{1}$. That is, we apply $K_{1}$ to $\theta$
to obtain $\theta^{*}$, then accept $\theta^{*}$ with probability
\begin{equation}
\alpha_{2}\left(\theta,\theta^{*}\right)=\min\left\{ 1,\frac{\pi_{2}\left(\theta^{*}\right)\pi_{1}\left(\theta\right)}{\pi_{2}\left(\theta\right)\pi_{1}\left(\theta^{*}\right)}\right\} .\label{eq:mh_within_mh}
\end{equation}
A computational saving arises when a rejection occurs under $K_1$, since in this case $\theta^* = \theta$, giving $\alpha_{2}=1$ without needing to evaluate $\pi_2$. Verifying this acceptance probability is simple. Since $\theta^{*}$
is simulated from $K_{1}$, we have an acceptance probability for
$K_{2}$ of

\begin{equation}
\alpha_{2}\left(\theta,\theta^{*}\right)=\min\left\{ 1,\frac{\pi_{2}\left(\theta^{*}\right)K_{1}\left(\theta\mid\theta^{*}\right)}{\pi_{2}\left(\theta\right)K_{1}\left(\theta^{*}\mid\theta\right)}\right\} .\label{e:mh_prop_for_mh_acceptance}
\end{equation}
Now, $K_{1}$ satisfies the detailed balance equation with respect
to $\pi_{1}$
\begin{equation}
\pi_{1}\left(\theta^{*}\right)K_{1}\left(\theta \mid\theta^{*} \right)=\pi_{1}\left(\theta\right)K_{1}\left(\theta^{*} \mid\theta \right),\label{e:mh_prop_for_mh_db}
\end{equation}
and substituting this into (\ref{e:mh_prop_for_mh_acceptance}) we obtain
the desired result of equation (\ref{eq:mh_within_mh}).

In the literature on delayed acceptance, $\pi_{2}$ is chosen to be
a desired target distribution that may be computationally expensive
to evaluate, and $\pi_{1}$ is chosen to be a~cheaper, approximate
target distribution. Used in this way, the first stage of delayed
acceptance (i.e. the result of applying $K_{1}$) provides a~proposal
that should be well suited for an MH algorithm with a~target
$\pi_{2}$, e.g. \citet{Banterle2014} uses approximate likelihoods
based on subsets of the data as the approximate targets. \citet{Strens2004}
uses a~very similar idea, i.e. a~hierarchical decomposition of the
likelihood. \citet{Banterle2014} note that whilst a~standard MH
algorithm dominates delayed acceptance in terms of the Peskun ordering,
it is possible that a~reduced computational cost in the first stage
of delayed acceptance can lead to more efficient algorithms in terms
of the Monte Carlo error per computational time.

\subsubsection{Delayed acceptance with ABC-MCMC\label{subsec:Delayed-acceptance-with}}

This paper makes use of delayed acceptance in the ABC setting. Suppose
that there exists a~computationally cheap alternative $l_{1}$ to
our ``true'' simulator $l_{2}=l$. We wish to perform the first stage
of delayed acceptance using points $x_{1}^{*}$ simulated from $l_{1}$, which are compared to data $y_1$ (which need not necessarily be the same as $y$) to determine the parameters at which we simulate points
$x_{2}^{*}$ from $l_{2}$, from which we estimate the standard ABC
likelihood. At the first stage of delayed acceptance, we use an ABC-MCMC
move with the simulator $l_{1}$, using a~tolerance $\epsilon_{1}$, giving an acceptance probability of

\begin{equation}
\alpha_{1} = \min\left\{ 1,\frac{p\left(\theta^{*}\right)P_{\epsilon_{1}}\left(y_{1}\mid x_{1}^{*}\right)}{p\left(\theta\right)P_{\epsilon_{1}}\left(y_{1}\mid x_{1}\right)}\frac{q\left(\theta\mid\theta^{*}\right)}{q\left(\theta^{*}\mid\theta\right)}\right\} .
\end{equation}
The acceptance probability at the second stage is
\[
\alpha_{2} = \min\left\{ 1,\frac{P_{\epsilon_{2}}\left(y\mid x_{1}^{*},x_{2}^{*}\right)}{P_{\epsilon_{2}}\left(y\mid x_{1},x_{2}\right)}\frac{P_{\epsilon_{1}}\left(y_{1}\mid x_{1}\right)}{P_{\epsilon_{1}}\left(y_{1}\mid x_{1}^{*}\right)}\right\},
\]
where the marginal distribution of the target we have used is the
ABC posterior with $l_{2}$, $\epsilon_2$ and $y$. Note that we have
included the conditioning on $x_{1}$ in $P_{\epsilon_{2}}$ and
$l_{2}$, which allows for the possibility that the ``true'' simulator
is a~composition of the simulators $l_{1}$ and $l_{2}$, with $x_{1}\sim l_{1}$
being a~partial simulation. If the true simulator is simply $l_{2}$,
we may drop this conditioning on $x_{1}$ from $P_{\epsilon_{2}}$
and $l_{2}$.

\subsection{Delayed acceptance ABC-SMC\label{subsec:Delayed-acceptance-ABC-SMC}}

\subsubsection{Use of DA-ABC-MCMC in ABC-SMC\label{subsec:Use-in-ABC-SMC}}


We now examine the use of the DA-ABC-MCMC
approach from the preceding section for the ``move'' step in the ABC-SMC algorithm
of \citet{DelMoral2012g}. The new SMC sampler operates on a~sequence of target distributions where $\epsilon_{2}$ decreases at each iteration, thus we use $\epsilon_{2,t}$
to denote its value at the $t$-th iteration. $\epsilon_{1}$ and $y_{1}$
may also change between SMC iterations, and we denote their values
at the $t$-th iteration by $\epsilon_{1,t}$ and $y_{1,t}$ respectively. In this case the weight update for each particle is given by
\begin{equation}
\tilde{w}_{t+1} = w_{t}\frac{P_{\epsilon_{2,,t+1}}\left(y\mid x_{1,t},x_{2,t}\right)}{P_{\epsilon_{2,,t}}\left(y\mid x_{1,t},x_{2,t}\right)}.
\end{equation}

\subsubsection{Adaptive DA within ABC-SMC when using the indicator kernel}

In this section we consider the implementation of the DA-ABC-MCMC move when $P_{\epsilon_{1,t}}$
is chosen to be an indicator function; i.e. $P_{\epsilon_{1,t}}\left(y_{1,t}\mid x_{1,t}^{*}\right)\propto\mathbb{I}\left(d\left(y_{1,t},x_{1,t}^{*}\right)<=\epsilon_{1,t}\right)$,
where $d$ is a~distance metric. Following section \ref{subsec:Delayed-acceptance-with} we obtain, at the $t$-th iteration of the SMC, an acceptance probabilities of
\begin{equation}
\alpha_{1,t}=\begin{cases}
\min\left\{ 1,\frac{p\left(\theta_{t}^{*}\right)}{p\left(\theta_{t}\right)}\frac{q\left(\theta_{t}\mid\theta_{t}^{*}\right)}{q\left(\theta_{t}^{*}\mid\theta_{t}\right)}\right\}  & \mbox{if } d\left(y_{1,t},x_{1,t}^{*}\right), d\left(y_{1,t},x_{1,t}\right) <\epsilon_{1,t}\\
0 & \mbox{otherwise.}
\end{cases}.\label{eq:1st_stage_acc_smc}
\end{equation}
and
\[
\alpha_{2,t}=\begin{cases}
1 & \mbox{if } d\left(y,x_{2,t}^{*}\right)<\epsilon_{2,t},\\
0 & \mbox{otherwise.}
\end{cases}
\]
at the first and second stages of DA respectively. The dependence in equation (\ref{eq:1st_stage_acc_smc}) on the condition $d\left(y_{1,t},x_{1,t}\right)$ is maybe surprising, but is required since if it is not satisfied the denominator in the acceptance ratio is zero, and in this case the proposal should be rejected \citep{Tierney1998}. The presence of this condition is due to the possibility that the sequence $\epsilon_{1,t}$ is not always decreasing in $t$. Full details may be found in the appendix, along with details of how early rejection may be used in the first stage of DA.

\citet{DelMoral2012g} make use of a~useful property of ABC to create an adaptive algorithm: once simulation of $x$ has been performed for any $\theta$,
it is computationally cheap to estimate the ABC likelihood for any
tolerance $\epsilon$. Here we use this property, together with the
fact that we have a~population of particles available in the SMC algorithm,
to automatically determine an appropriate value for $\epsilon_{1,t}$
at every iteration. We choose $\epsilon_{1,t}$ using the criterion
that we desire to perform the second stage of the delayed acceptance
for a~fixed proportion of the particles; we choose $\epsilon_{1,t}$
such that $A$ particles are accepted to move forward to the second
stage, where $0<A \leq N$
is chosen prior to running the SMC. This is achieved by choosing $\epsilon_{1,t}$
such that $A$ particles satisfy $d\left(y_{1,t},x_{1,t}\right),d\left(y_{1,t},x_{1,t}^{*}\right)<\epsilon_{1,t}$:
a bisection method may be used to find such an $\epsilon_{1,t}$. As with the bisection used in the adaptive approach of \citet{DelMoral2012g}, in practice the bisection will not always give precisely $A$ particles at the second stage of DA; if fewer than $A$ particles pass early rejection due to the prior, $\epsilon_{1,t}$ is chosen to let all these particles through to the second stage.

\subsubsection{Discussion}

This ABC-SMC algorithm, which we call DA-ABC-SMC, is described in full in algorithm \ref{alg:da_abc_smc}. In brief, we use this algorithm where we: adaptively choose the sequence $\left( \epsilon_{2,t} \right)$ such that there are $U$ unique particles after reweighting and resampling \citep{Bernton2017}; adaptively choose the variance of the (Gaussian) MCMC proposals to be the sample variance of the current particle set. The method requires us to specify: the number of particles $N$, the
number of particles $0<A\leq N$ that are allowed past the first stage of
the DA and the number of unique particles $0<U\leq N$. We now discuss how to choose each
of these parameters. 
\begin{itemize}
\item \textbf{Number of unique particles. }On terminating the algorithm,
we effectively have a~sample of size $U$ from the final ABC posterior,
thus $U$ should be chosen to be the number of Monte Carlo points
we wish to generate.
\item \textbf{Number of particles allowed past the first stage. }$A$ dictates
the computational cost of each iteration, since it is the number of
expensive simulations we will perform per SMC iteration.
\item \textbf{Total number of particles.} To choose $N$ we need to
consider the factor $F$ by which the cheap simulator is faster than
the expensive simulator. For DA to be most effective, we require that
the cost of running the cheap simulator is small compared to running
the expensive simulator, i.e. $N/F\ll A$.
\end{itemize}
If $N$ is chosen to be much larger than $A$, we require a~slightly
non-standard initial step in our SMC algorithm so that the computational
expense of this step does not dominate the subsequent iterations.
In a~standard SMC sampler, the initial step would require initial
values of $x_{1}$ and $x_{2}$ to be simulated for each of the $N$
particles, which requires running the expensive simulator $N$ times.
Instead we simulate $\left(\theta,x_{1},x_{2}\right)$ $A$ times
from the initial distribution, and repeat these points $N/A$ times (i.e. the vector of particles consists of stacked copies of these values)
so that we have a~sample of size $N$ from the initial distribution.
This sample contains only $A$ unique points but, importantly, will
result in $N$ proposed points at every MCMC move in the SMC.
The parameter $N$ plays slightly different role to a~standard
SMC sampler, in which we might informally think of it as roughly the
size of the Monte Carlo sample (in DA-ABC-SMC, $U$ plays this role
instead). In DA-ABC-SMC, $N$ is one of four factors determining
the ``DA proposal'' (i.e. the distribution of points that arrive
at the second stage of DA), the remaining factors being: the choices
of $A$ and $l_{1}$, and the distribution of the particles from the previous
iteration of the SMC. A~useful DA proposal has similar characteristics
to a~good independent MCMC or importance sampling proposal: we wish
it to be close to the target distribution, with wider tails. Since
the DA proposal depends on the previous distribution of the particles,
we observe empirically that it can to some extent automatically ``track'' the target
distribution over SMC iterations, and provide a~useful proposal distribution
at each SMC iteration. The extent to which this is the case is investigated in sections \ref{sec:Application-to-stochastic} and \ref{sec:Application-to-Ising}, where we illustrate the performance of our algorithm for different choices of the tuning parameters, which result in different sequences of DA proposals.

\begin{algorithm}[H]
\caption{DA-ABC-SMC using early rejection.}
\label{alg:da_abc_smc}
\hspace*{\algorithmicindent} \textbf{Inputs:} As algorithm \ref{alg:abc_smc}, and the number of particles $A$ to pass the first stage of DA. \\
 \hspace*{\algorithmicindent} \textbf{Outputs:} Particles $\left\{\left( \theta^{(i)}_t,x^{(i)}_t \right) \right\}_{i=1}^N$ and weights $\left\{w_{t}^{(i)} \right\}_{i=1}^N$ for all $t$.
\begin{algorithmic}
\For {$i= 1:A$}
    \State $w^{((i-1)N/A+1:iN/A)}_0 = 1/N$, $\theta^{((i-1)N/A+1:iN/A)}_0 \sim p\left( \cdot \right)$
    \State $x^{((i-1)N/A+1:iN/A)}_{1,0} \sim l_1\left( \cdot \mid \theta^{(iN/A)}_0 \right)$, $x^{((i-1)N/A+1:iN/A)}_{2,0} \sim l_2\left( \cdot \mid \theta^{(iN/A)}_0, x^{((iN/A)}_{1,0} \right)$
\EndFor
\State $\epsilon_{1,0} = \epsilon_{1,\mbox{start}}$, $\epsilon_{2,0} = \epsilon_{2,\mbox{start}}$, $t=0$.
\While {$\epsilon_{2,t} > \epsilon_{2,\mbox{end}}$}
    \State Simulate $v \sim \mathcal{U}\left[ 0,1 \right]^N$, to be used in resampling and use bisection to choose $\epsilon_{2,t+1}$ s.t. there will be $U$ unique particles after reweighting and resampling (using $v$).
    \For {$i= 1:N$}
        \State $\tilde{w}_{t+1}^{(i)}=w_{t}^{(i)}\mathbb{I}\left(d\left(y,x_{2,t}^{(i)}\right)<\epsilon_{2,t+1}\right)$
    \EndFor
    \State Normalise $\left\{ \tilde{w}_{t+1} \right\}_{i=1}^N$ to give normalised weights $\left\{ w_{t+1} \right\}_{i=1}^N$, perform resampling using random numbers $v$, and let $I = \emptyset$.
    \For {$i= 1:N$}
        \State $\theta^{(i)}_{t+1} =  \theta^{(i)}_{t}$, $x^{(i)}_{1,t+1} = x^{(i)}_{1,t}$, $x^{(i)}_{2,t+1} =  x^{(i)}_{2,t}$
        \State $\left(\theta^{(i)}_{t+1}\right)^* \sim q\left( \cdot \mid \theta^{(i)}_{t} \right)$
        \State $u \sim \mathcal{U}\left(0,1\right)$
        \If {$u < \frac{p\left( \left(\theta^{(i)}_{t+1}\right)^* \right) q\left( \theta^{(i)}_{t} \mid \left(\theta^{(i)}_{t+1}\right)^* \right) }{ p\left( \theta^{(i)}_{t+1} \right) q\left( \left(\theta^{(i)}_{t+1}\right)^* \mid \theta^{(i)}_{t} \right) }$ }
            \State $ \left(x^{(i)}_{1,t+1} \right)^* \sim l_1\left( \cdot \mid \left(\theta^{(i)}_{t+1}\right)^* \right)$
        \EndIf
    \EndFor
    \State Use bisection to choose $\epsilon_{1,t+1}$ s.t. $A$ particles satisfy $d\left(y_{1,t},x_{1,t}\right),d\left(y_{1,t},x_{1,t}^{*}\right)<\epsilon_{1,t}$.
    \State Let $I$ be the set of indices of the particles that pass the first stage of DA.
    \For {$i \in I$}
        \State $ \left(x^{(i)}_{2,t+1} \right)^* \sim l_1\left( \cdot \mid \left(\theta^{(i)}_{2,t+1}\right)^*, \left(x^{(i)}_{1,t+1} \right)^* \right)$
        \If { $d\left(y, \left(x^{(i)}_{2,t+1} \right)^* \right) < \epsilon_{2,t+1}$ }
            \State $\theta^{(i)}_{t+1} =  \left(\theta^{(i)}_{t+1}\right)^*$, $x^{(i)}_{1,t+1} =  \left(x^{(i)}_{1,t+1}\right)^*$, $x^{(i)}_{2,t+1} =  \left(x^{(i)}_{2,t+1}\right)^*$
        \EndIf
    \EndFor
    \State $t = t + 1$
\EndWhile
\end{algorithmic}
\end{algorithm}


\section{Application to SDEs\label{sec:Application-to-stochastic}}

\subsection{Lotka-Volterra model}

The Lotka-Volterra model is a~stochastic Markov jump process that
models the number of individuals in two populations of animals: predator and prey. It is a~commonly
used example in ABC since it is possible to simulate exactly from
the model using the Gillespie algorithm \citep{Gillespie1977}, but
the likelihood is not available pointwise. We follow the model described in \citet{Wilkinson2011}, in which $X$ represents the number
of predators and $Y$ the number of prey. The following reactions
may take place: a prey may be born, with rate $\theta_{1}Y$, increasing $Y$ by one; predator and prey may interact, with rate $\theta_{2}XY$, increasing $X$ by one and decreasing $Y$ by one; a~predator may die, with rate $\theta_{3}X$, decreasing $X$ by one.
\citet{Papamakarios2016} note that for most parameters the size of one population quickly decreases
to zero (in the case of the predators dying out, this results in the
prey population growing exponentially). The relatively small region
of parameter space that contains parameter values resulting
in oscillating population sizes makes this a~relatively challenging inference problem.

In this section we use this example to demonstrate the use of DA-ABC-SMC
for SDE models that need to be simulated numerically. To do this,
we use the chemical Langevin equation approximation to the Markov
jump process, as detailed in \citet{Golightly2015}. This results
in two coupled non-linear SDEs, which we simulate numerically using
the Euler-Maruyama method.

\subsection{Results}

All of our empirical results were generated using R \citep{R2019}, and the R packages \texttt{ggplot2} \citep{ggplot2}, \texttt{matlab} \citep{matlab} and \texttt{mvtnorm} \citep{mvtnorm} were used. We study the data ``\texttt{LVPerfect}'' in the R package \texttt{smfsb} \citep{smfsb} (the numerical methods for simulating from the likelihood are also taken from this
package). This data, which was generated using parameter values $\left(\theta_1 = 1, \theta_2 = 0.05, \theta_3 = 0.6 \right)$, has been previously studied in \citet{Wilkinson2011}, and we use the same priors and ABC approach as in this reference. We used DA-ABC-SMC with a variety of choices of $U$, $A$ and $N$, and two different choices of the Euler-Maruyama step size $s$ in
the cheap simulator $s=0.5$ and $s=0.1$, both of which result in very rough approximations of the dynamics.  We compared these approaches with standard ABC-SMC,
with $N=200$ particles and a~sequence of tolerances selected
such that $U=100$ unique particles are retained at each iteration, an SMC$^2$-style approach \citep{Duan2015} (that uses $M$ particles when using a particle filter to estimate the likelihood) and also, as black horizontal lines, ``ground truth'' for the posterior expectation and standard deviation of the parameters found using a long run ($10^5$ iterations) of ABC-MCMC. All algorithms we run 30 times, and measure computational cost, we counted
the total number of steps $\mathbf{S}$ (taking the median $\bar{\mathbf{S}}$ over the 30 runs) simulated using Euler-Maruyama. The appendix contains the full details.

\begin{figure}[H]
\centering
\begin{subfigure}{.3\textwidth}
  \centering
  \includegraphics[width=1\linewidth]{{{mean1_v_sims_5e-04_0.1_500_100_0.2_0.15_}}}
  \caption{DA-ABC-SMC: $N=500$, $U=100$, $A=100$, $s=0.1$.}
\end{subfigure}%
\hspace{0.5cm}
\begin{subfigure}{.3\textwidth}
  \centering
  \includegraphics[width=1\linewidth]{{{mean1_v_sims_5e-04_0.1_5000_100_0.02_0.15_}}}
  \caption{DA-ABC-SMC: $N=5000$, $U=100$, $A=100$, $s=0.1$.}
\end{subfigure}%
\hspace{0.5cm}
\begin{subfigure}{.3\textwidth}
  \centering
  \includegraphics[width=1\linewidth]{{{mean1_v_sims_5e-04_0.1_200_100_0.5_0.15_}}}
  \caption{ABC-SMC: $N=200$, $U=100$.}
\end{subfigure}
\caption{The estimated posterior mean plotted against the total number of time steps used in Euler-Maruyama.} \label{f:lv_post_mean}
\end{figure}

\begin{figure}[H]
\centering
\begin{subfigure}{.3\textwidth}
  \centering
  \includegraphics[width=1\linewidth]{{{sd1_v_sims_5e-04_0.1_500_100_0.2_0.15_}}}
  \caption{DA-ABC-SMC: $N=500$, $U=100$, $A=100$, $s=0.1$.}
\end{subfigure}%
\hspace{0.5cm}
\begin{subfigure}{.3\textwidth}
  \centering
  \includegraphics[width=1\linewidth]{{{sd1_v_sims_5e-04_0.1_5000_100_0.02_0.15_}}}
  \caption{DA-ABC-SMC: $N=5000$, $U=100$, $A=100$, $s=0.1$.}
\end{subfigure}%
\hspace{0.5cm}
\begin{subfigure}{.3\textwidth}
  \centering
  \includegraphics[width=1\linewidth]{{{sd1_v_sims_5e-04_0.1_200_100_0.5_0.15_}}}
  \caption{ABC-SMC: $N=200$, $U=100$.}
\end{subfigure}
\caption{The estimated posterior standard deviation plotted against the total number of time steps used in Euler-Maruyama.} \label{f:lv_post_sd}
\end{figure}

Figures \ref{f:lv_post_mean} and \ref{f:lv_post_sd} show, for every run of three methods, the evolution of the sample mean and standard deviation of the particles against the number of time steps of the numerical solver, with the final red dot indicating the estimate from the final distribution (where $\epsilon_{2}=0.15$ was reached). The black line on each plots shows estimated ground truth values estimated using 20 runs of ABC-MCMC of length 5,000 each. We only show the results for $\theta_{1}$ here, since the results for the other parameters are comparable in terms of how they illustrate the properties of the algorithms. The method on the left is DA-ABC-SMC with $N=500$ and $U=A=100$; in the middle is DA-ABC-SMC with $N=5000$ and $U=A=100$; and on the right is ABC-SMC with $N=200$ and $U=100$. We observe that whilst ABC-SMC takes significantly more steps to converge to the target distribution, the estimates of the mean and standard deviation have a relatively low variance. The DA-ABC-SMC approaches converge much faster, but result in estimates that are usually further from the ground truth. In addition, both DA-ABC-SMC approaches appear to underestimate the posterior standard deviation, with this effect becoming more pronounced when $N$ is larger. This result suggests that choosing $N$ too large (resulting in $A/N$ being small) can result in a DA proposal that is too concentrated compared to the target. To further elaborate on this important point, as the ratio $A/N$ decreases, the first stage of the DA leads to a very poor proposal - this is because the particles that make it through to the second stage of the DA step are those that have simulations that result in the very smallest distance to the observed data. The effect is to make the DA proposal too concentrated around the mode of the posterior yielded by the fast approximation.

Tables \ref{t:mean} and \ref{t:sd} summarise the properties of the estimates of the posterior mean and standard deviation from each method. The bias, standard deviation and root mean square error (RMSE) are reported, along with the RMSE$\times \sqrt{\bar{\mathbf{S}}}$, to help compare the errors of methods with different computational costs. The results for some configurations are heavily influenced by poor results from a single run, for example the case of DA-ABC-SMC with $N=1000$, $U=100$, $A=100$ and $s=0.1$. However, based on these tables and the plots of the estimated posterior mean and standard deviation in the appendix, there is evidence to make the following conclusions.

\begin{table}[ht] \label{t:param1}
\centering
\begin{tabular}{rrrrrrrrr}
  \hline
Method & Bias & Std. dev. & RMSE & RMSE$\times \sqrt{\bar{\mathbf{S}}}$ \\ 
  \hline
  DA-ABC-SMC: $N=500$, $U=100$, $A=100$, $s=0.1$. & 0.0033 & 0.0625 & 0.0626 & 2235  \\ 
  DA-ABC-SMC: $N=500$, $U=100$, $A=100$, $s=0.5$. & -0.0047 & 0.0439 & 0.0442 & 1449 \\ 
  DA-ABC-SMC: $N=1000$, $U=100$, $A=100$, $s=0.1$. & 0.0659 & 0.2294 & 0.2387 & 6861 \\ 
  DA-ABC-SMC: $N=1000$, $U=100$, $A=100$, $s=0.5$. & -0.0003 & 0.0381 & 0.0381 & 1012 \\ 
  DA-ABC-SMC: $N=1000$, $U=200$, $A=100$, $s=0.1$.& 0.0006 & 0.0559 & 0.0559 & 4320  \\ 
  DA-ABC-SMC: $N=1000$, $U=50$, $A=100$, $s=0.1$. & -0.0046 & 0.0754 & 0.0756 & 1232 \\ 
  DA-ABC-SMC: $N=5000$, $U=100$, $A=100$, $s=0.1$. & -0.0097 & 0.1094 & 0.1099 & 2401 \\ 
  DA-ABC-SMC: $N=10000$, $U=100$, $A=100$, $s=0.1$. & -0.0459 & 0.0703 & 0.0840 & 1960 \\ 
  SMC$^2$: $N=100$, $M=100$, $\alpha=0.9$, $s=0.1$. & 0.2929 & 1.2136 & 1.2485 & 35065 \\ 
  SMC$^2$: $N=100$, $M=100$, $\alpha=0.99$, $s=0.1$. & 0.3261 & 1.0518 & 1.1012 & 55438 \\ 
  SMC$^2$: $N=100$, $M=1000$, $\alpha=0.9$, $s=0.1$. & 0.9847 & 1.6379 & 1.9111 & 150838 \\ 
  ABC-SMC: $N=200$, $U=100$. & -0.0092 & 0.0692 & 0.0698 & 3718 \\ 
  ABC-SMC*: $N=200$, $U=100$. & 0.0081 & 0.0583 & 0.0589 & 3136 \\ 
   \hline
\end{tabular}
\caption{Properties of estimators of the posterior mean of $\theta_1$. *The final line of the table refers to ABC-SMC where the runs that did not terminate are excluded from the calculation of the statistics.} \label{t:mean}
\end{table}

\begin{table}[ht] \label{t:param1}
\centering
\begin{tabular}{rrrrrrrrr}
  \hline
Method & Bias & Std. dev. & RMSE & RMSE$\times \sqrt{\bar{\mathbf{S}}}$ \\ 
  \hline
  DA-ABC-SMC: $N=500$, $U=100$, $A=100$, $s=0.1$. & -0.0178 & 0.0299 & 0.0348 & 1242 \\ 
  DA-ABC-SMC: $N=500$, $U=100$, $A=100$, $s=0.5$. & -0.0403 & 0.0200 & 0.0450 & 1476 \\ 
  DA-ABC-SMC: $N=1000$, $U=100$, $A=100$, $s=0.1$. & 0.0207 & 0.2368 & 0.2377 & 6833 \\ 
  DA-ABC-SMC: $N=1000$, $U=100$, $A=100$, $s=0.5$. & -0.0488 & 0.0194 & 0.0525 & 1393 \\ 
  DA-ABC-SMC: $N=1000$, $U=200$, $A=100$, $s=0.1$. & -0.0104 & 0.0275 & 0.0294 & 2273 \\ 
  DA-ABC-SMC: $N=1000$, $U=50$, $A=100$, $s=0.1$. & -0.0507 & 0.0277 & 0.0578 & 942 \\ 
  DA-ABC-SMC: $N=5000$, $U=100$, $A=100$, $s=0.1$. & -0.0568 & 0.0307 & 0.0645 & 1410 \\ 
  DA-ABC-SMC: $N=10000$, $U=100$, $A=100$, $s=0.1$. & -0.0621 & 0.0277 & 0.0680 & 1587 \\ 
  ABC-SMC: $N=200$, $U=100$. & 0.2321 & 0.5136 & 0.5636 & 30023 \\ 
  ABC-SMC*: $N=200$, $U=100$. & 0.0175 & 0.1105 & 0.1119 & 5961 \\ 
   \hline
\end{tabular}
\caption{Properties of estimators of the posterior standard deviation of $\theta_1$. *The final line of the table refers to ABC-SMC where the runs that did not terminate are excluded from the calculation of the statistics.} \label{t:sd}
\end{table}

\begin{itemize}

\item Choosing $N$ too large (resulting in $A/N$ being small) can result in a DA proposal that is too concentrated compared to the target, and leads to underestimating the posterior standard deviation.

\item Even when $N$ is not large, in this example often DA-ABC-SMC results in a sample that underestimates the standard deviation of the posterior distribution. The likely cause of this is that the DA proposal is too concentrated. This proposal easily generates unique points, thus the tolerance $\epsilon_2$ is reduced too quickly, resulting in a poor quality sample. The results from DA-ABC-SMC when $N=1000$, $U=200$, $A=100$ and $s=0.1$, where 200 instead of 100 unique points are required when reducing the tolerance, suggests that when the SMC does not reduce the tolerance as fast (and hence more MCMC moves are made), the estimates of the posterior standard deviation are more accurate.

\item DA-ABC-SMC can offer improved performance over ABC-SMC when measured by RMSE scaled by computational cost. This is largely due to its ability to quickly locate the region of highest posterior mass. However, it may not provide a representative sample from the posterior if the DA proposal is not tuned well. In contrast, compared to DA-ABC-SMC, standard ABC-SMC may be extremely slow in converging at all.

\item The SMC$^2$ approaches tried here do not lead to an improvement over the ABC approaches, due to the low acceptance rate of the particle MCMC moves used in the move step of the SMC sampler (a consequence of the high variance likelihood estimates). In many cases no proposals were accepted at an SMC iteration, leading to a poor quality sample.

\end{itemize}

\section{Applications to doubly intractable distributions\label{sec:Application-to-Ising}}

\subsection{Background}

This section concerns the class of likelihoods that may not
be evaluated pointwise due to the presence of an~intractable normalising
constant, i.e.
\[
l\left(y\mid\theta\right)=\frac{\gamma\left(y\mid\theta\right)}{Z\left(\theta\right)},
\]
where $\gamma\left(y\mid\theta\right)$ is tractable, but it is not
computationally feasible to evaluate the normalising constant $Z\left(\theta\right)$
(also known as the \emph{partition function}). Such a~distribution
is known as \emph{doubly intractable} \citep{Murray2006} since it
is not possible to directly use the Metropolis-Hastings algorithm
to simulate from the posterior $\pi\left(\theta\mid y\right)$, due
to the presence of $Z\left(\cdot\right)$ in both the numerator and
denominator of the acceptance ratio. The most common occurrence of
these distributions occurs when $l$ factorises as a~Markov random
field. The most well-established approaches to inference for these
models are the single and multiple auxiliary variable approaches \citep{Moller2006,Murray2006}
and the exchange algorithm \citep{Murray2006}, which respectively
use importance sampling estimates of the likelihood and acceptance
ratio to avoid the calculation of the normalising constant. From here
on we refer to these as ``auxiliary variable methods''. ABC \citep{Grelaud2009}
and synthetic likelihood \citep{Everitt2017} have also been used
for inference in these models. See \citet{Stoehr2017} for a~recent,
thorough review of the literature.

Previous work, e.g. \citet{Friel2013e}, suggests that
auxiliary variable methods are more effective than ABC for simulating
from the posterior $\pi\left(\theta\mid y\right)$ when $l$ has
an intractable normalising constant. In the full data ABC case, \citet{Everitt2017c}
suggest that this is because the multiple auxiliary variable approach
may be seen to be a~carefully designed non-standard ABC method. However,
in this paper we consider the situation in \citet{Everitt2012}, where
our data $y$ are indirect observations of a~latent (hidden) field
$x^{h}$, modelled with a~joint distribution $p\left(\theta\right)l\left(x^{h}\mid\theta\right)g\left(y\mid x^{h},\theta\right)$
with $l\left(x^{h}\mid\theta\right)$ having an intractable normalising
constant. In this case we might expect ABC to become more competitive,
or even to outperform other approaches. The most obvious approach is
to use data augmentation: using MCMC to simulate from the joint posterior
$\pi\left(\theta,x^{h}\mid y\right)$ by alternating simulation from
the full conditionals of $\theta$ and $x^{h}$, with the exchange
algorithm in the $\theta$ update. \citet{Everitt2012} compares this
approach with ``exchange marginal particle MCMC'', in which an SMC
algorithm is used to integrate out the $x^{h}$ variables at every
iteration of an MCMC and finds the particle MCMC approach to be preferable.
However, in order to perform well, both of these approaches require
efficient simulation from the conditional distribution $\pi\left(x^{h}\mid\theta,y\right)$.

Now compare these ``exact'' approaches with ABC. In this case, for
every $\theta$ we simulate $x^{h}\sim l\left(\cdot\mid\theta\right)$
and $x\sim g\left(\cdot\mid x^{h},\theta\right)$, then use an ABC
likelihood that compares $S\left(x\right)$ with $S\left(y\right)$.
The $x^{h}$ variables corresponding to a~particular $\theta$ variable
that are retained in the ABC sample are distributed according to an
ABC approximation to $\pi\left(x^{h}\mid\theta,S\left(y\right)\right)$;
indeed, in order that the ABC is at all efficient, there must be a
reasonable probability that simulated $x^{h}$ is in a~region of high
probability mass under this distribution. Compared to particle MCMC:
\begin{itemize}
\item Standard ABC has the disadvantage that the simulation of $x^{h}$
is, in contrast to particle MCMC, not conditioned on $y$ (although
we note recent work in \citet{Prangle2016} in which for some models
we may also refine the ABC likelihood by conditioning on $y$).
\item ABC has the advantage that \emph{a posteriori} $x^{h}$ is only conditioned
on some statistic $S\left(y\right)$, rather than $y$ as in particle MCMC. This
condition is often considerably less stringent. For example, consider
the Ising model example described below. When conditioning on $y$,
the posterior $\pi\left(x^{h}\mid\theta,y\right)$ is often restricted
to relatively small regions of $x^{h}$-space: for individual pixel
of $x^{h}$ the posterior mass may be concentrated in a~small region
in order to match each individual data point. However, when conditioning
on $S\left(y\right)$, the posterior $\pi\left(x^{h}\mid\theta,S\left(y\right)\right)$
may have non-negligible mass in many regions of $x^{h}$-space: there
are many different configurations of pixels that give rise to similar
summary statistics.
\end{itemize}
In this paper we revisit using ABC on these models, examining the data
previously studied in \citet{Everitt2012}, and show it is possible to make computational savings using DA-ABC-SMC. We now introduce
the cheap simulator that makes this improvement possible. In Markov
random field models, exact simulation from $l\left(\cdot\mid\theta\right)$
is either very expensive or intractable. \citet{Grelaud2009} proposes
to replace the exact simulator with taking the last point of a~long
MCMC run, and \citet{Everitt2012} shows that, under certain ergodicity
conditions, as the length of this ``internal'' MCMC goes to infinity,
the bias in our ABC posterior sample due to this particular approximation
also goes to zero (following ideas in \citet{Andrieu2009}). However,
empirically one finds that often using only a~single MCMC iteration
is sufficient to yield a~posterior close to the desired posterior.
In this paper we propose to use an internal MCMC run of a~single iteration
as the cheap simulator, and to run the internal MCMC for many iterations
(we use 1,000) as the expensive simulator, where the final iteration of
the cheap simulator is used as the first iteration of the expensive
simulator. The next section presents an application to latent Ising models, with an application to ERGMs in the appendix.

\subsection{Latent Ising model\label{subsec:Latent-Ising-model}}

An Ising model is a~pairwise Markov random field model on binary variables,
each taking values in $\left\{ -1,1\right\} $. Its distribution is
given by
\[
l\left(x^{h}\mid\theta_{x}\right)\propto\exp\left(\theta_{x}\sum_{\left(i,j\right)\in\mathbf{N}}x_{i}^{h}x_{j}^{h}\right),
\]
where $\theta_{x}\in\mathbb{R}$, $x_{i}^{h}$ denotes the $i$-th
random variable in $x^{h}$ and $\mathbf{N}$ is a~set defining pairs of nodes that are \textquotedblleft neighbours\textquotedblright .
We consider the case where the neighbourhood structure is given by
a regular 2-dimensional grid. Our data $y$ are noisy observations
of this $x^{h}$ field: the $i$-th variable in $y$ has distribution
\[
g\left(y_{i}\mid x_{i}^{h},\theta_{y}\right)\propto\exp\left(\theta_{y}y_{i}x_{i}^{h}\right),
\]
as in \citet{Everitt2012} (with the normalising constant being intractable).
We used independent Gaussian priors on $\theta_{x}$ and $\theta_{y}$,
each with mean 0 and standard deviation 5. Our DA-ABC-SMC algorithm
used $U=A=100$, and we examined different values of $N$. A~single
site Gibbs sampler was used to simulate from the likelihood, with
the expensive simulator using the final point of 1,000 sweeps of the
Gibbs sampler. The MCMC proposal was taken to be a~Gaussian distribution
centred at the current particle, with covariance given by the sample
covariance of the particles from the previous iteration. We used a
single statistic in the ABC: the number of equivalently valued neighbours
$S\left(y\right)=\sum_{\left(i,j\right)\in\mathbf{N}}y_{i}y_{j}$
(noting that this is not sufficient, but that the particle MCMC results
in \citet{Everitt2012} indicate that this does not have a~large impact
on the posterior). The distance metric used in ABC on the summary
statistics was the absolute value of the difference.

\begin{figure}
\centering
\begin{subfigure}{.5\textwidth}
  \centering
  \includegraphics[width=1\linewidth]{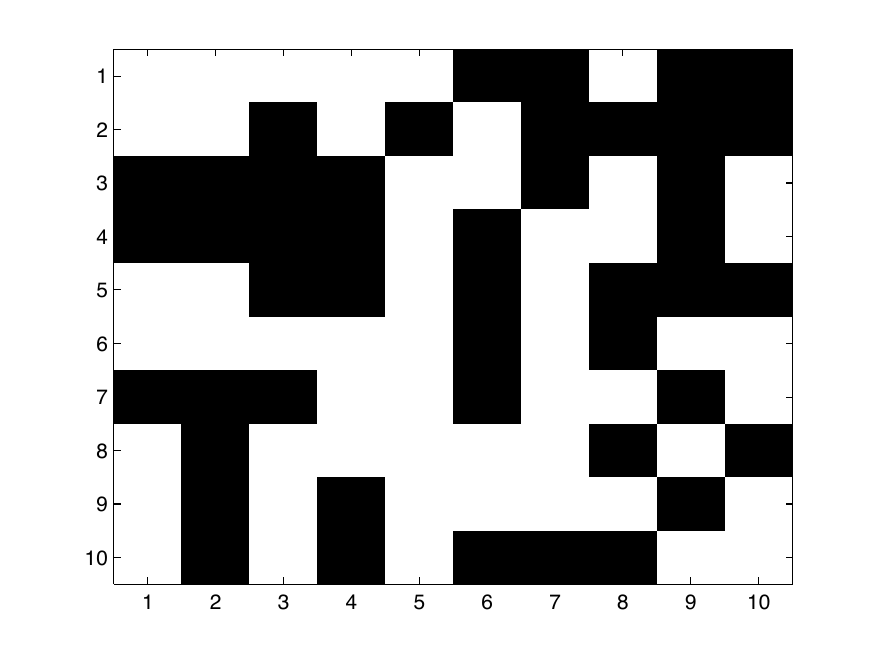}
  \caption{Observed data.}
  \label{fig:Observed-data.}
\end{subfigure}%
\begin{subfigure}{.5\textwidth}
  \centering
  \includegraphics[width=1\linewidth]{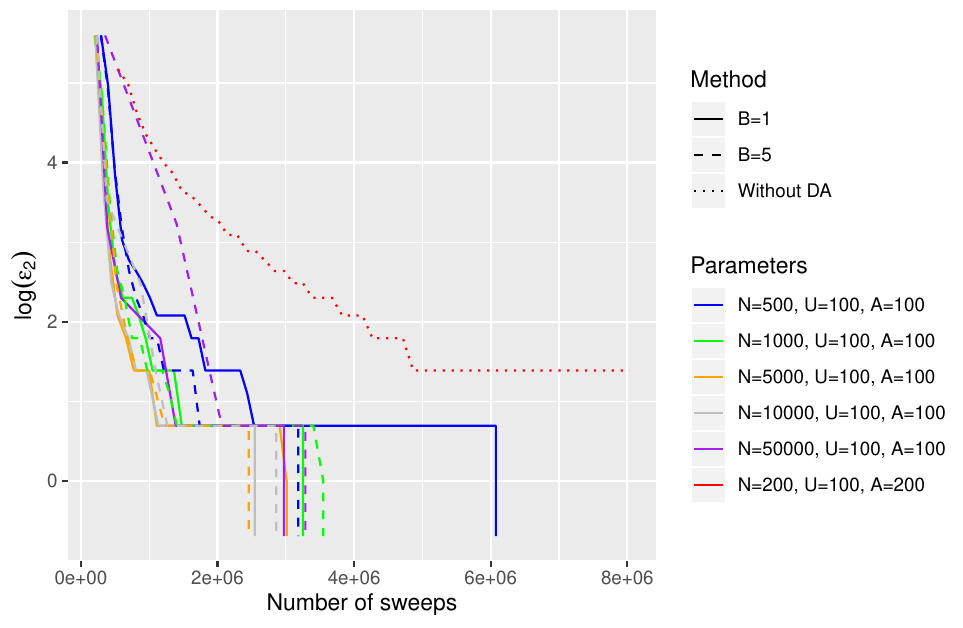}
  \caption{$\log\left(\epsilon_{2,t}\right)$ plotted against the number of sweeps
of the Gibbs sampler. When $\epsilon_{2,t}=0$ we plot $\log\left(\epsilon_{2,t}\right)=-\log\left(2\right)$.}
  \label{fig:-plotted-against}
\end{subfigure}
\begin{subfigure}{.5\textwidth}
  \centering
  \includegraphics[width=1.6\linewidth]{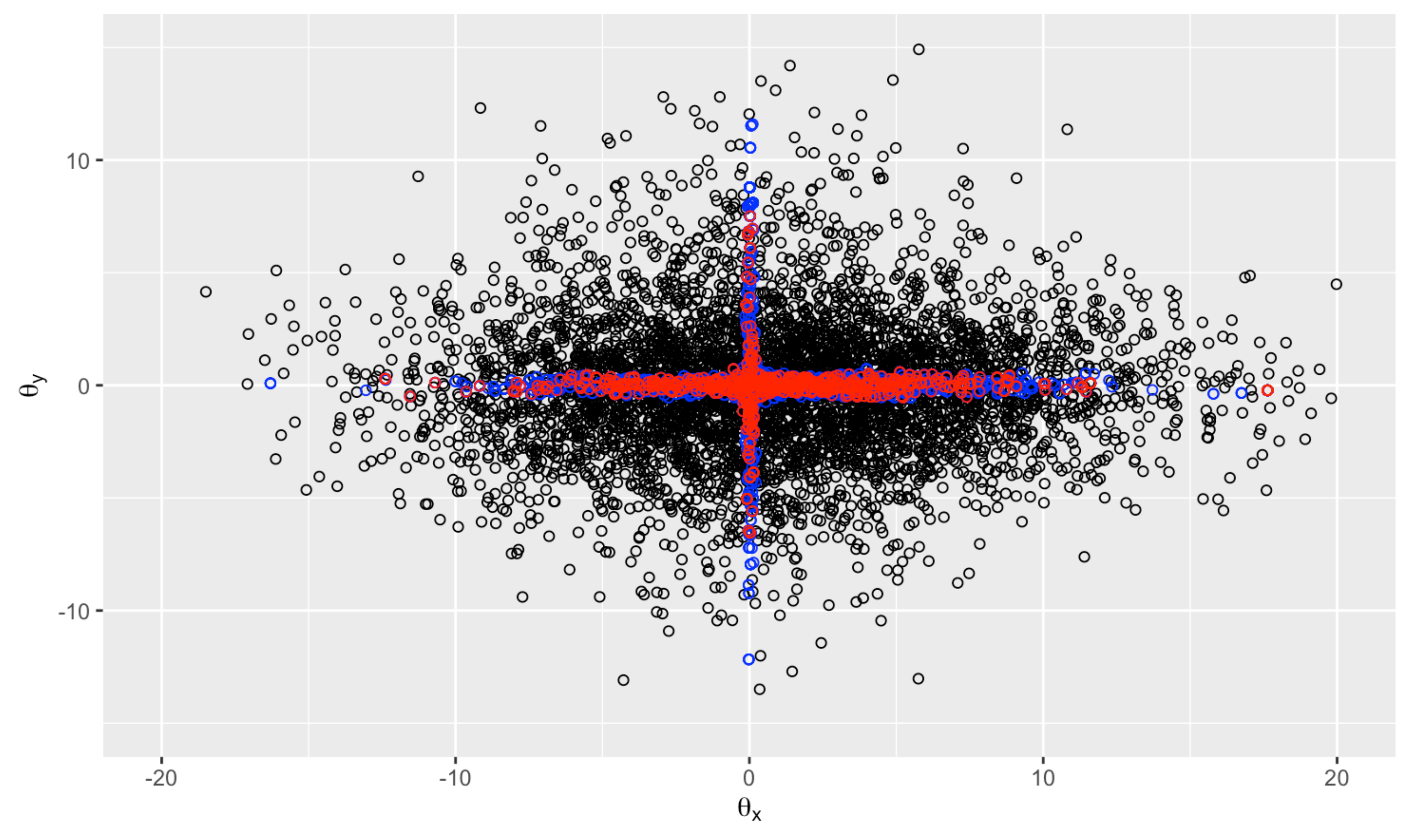}
  \caption{Points drawn at a~single iteration of DA-ABC-SMC with $N=5000$, $A=U=1000$. Proposed points (black) are plotted alongside points from the DA proposal (red), with $\epsilon_{1}=68$, and points from the posterior (blue), with $\epsilon_{2}=8$.}
  \label{fig:DA-proposal-when}
\end{subfigure}
\caption{DA-ABC-SMC applied to the latent Ising model.}
\end{figure}

We used the same $10\times10$ grid of data as was studied in \citet{Everitt2012},
shown in figure \ref{fig:Observed-data.}, which was generated from
the model using $\theta_{x}=\theta_{y}=0.1$. These parameter values
represent a~relatively weak interaction between neighbouring pixels,
and also quite noisy observations. On a~grid of this size, there is
ambiguity as to whether this data may have been generated with either
one or both of $\theta_{x}$ and $\theta_{y}$ being small, giving
a posterior distribution shaped like a~cross, similar to that shown in figure \ref{fig:DA-proposal-when}. Figure \ref{fig:DA-proposal-when} gives an example
of a~DA proposal, i.e. points that pass the first stage of the DA-MCMC,
from the early stages of a~run of DA-ABC-SMC. We observe how this
distribution would be a~suitable independent MCMC proposal for the
posterior.

All of our empirical results were generated using R \citep{R2019}, and the R packages \texttt{ggplot2} \citep{ggplot2}, \texttt{matlab} \citep{matlab} and \texttt{mvtnorm} \citep{mvtnorm} were used. We ran DA-ABC-SMC for several values of $N$, and compared the results
with standard ABC-SMC using all of the same algorithmic choices, with
$N=200$ and $U=100$. The standard ABC-SMC algorithm used the expensive
simulator, i.e. 1,000 iterations of the internal Gibbs sampler. In
algorithms we terminated the method when $\epsilon_{2}=0$, or after 500 SMC iterations,
whichever was sooner. We focus on studying the total computational
effort expended in each algorithm, measured by the total number of
sweeps $\mathbf{S}$ of the internal Gibbs sampler, and on the mean and standard deviation of the Euclidean distance $\rho$ of the parameter vector $(\theta_{x},\theta_{y})$ from the origin.

We ran the algorithm with several different values of $N$, and two
different cheap simulators: the first where only a~single sweep of
the Gibbs sampler is used ($B=1$); the second using the final point
of 5 sweeps of the Gibbs sampler ($B=5$). Each configuration was run 30 times. We chose $N$ such that
in most cases the computational cost was dominated by the expensive
simulations in the second stage of the DA, although when $N=50,000$
and $B=5$ the cost of the first stage dominated. Figure \ref{fig:-plotted-against}
shows gives an example showing the evolution of $\epsilon_{2}$ against the total number of sweeps of the Gibbs sampler for several different configurations. We see that all configurations of the DA approach appear to have a significantly lower cost than standard ABC-SMC: in all cases the DA approach reaches
$\epsilon_{2}=0$ whereas the ABC-SMC reaches $\epsilon_{2}=2$.

Table \ref{t:rho} shows properties of estimates of the mean and standard deviation posterior on $\rho$. The estimates from ABC-SMC are based on the output of the 500th iteration of the SMC, where in all runs $\epsilon_{2}=2$. we observe that we obtain lower variance estimates from ABC-SMC than from DA-ABC-SMC, but at a higher cost. This cost may be significantly higher if we wish to reduce $\epsilon_2$ to 0, as is achieved in all cases by DA-ABC-SMC. As in the previous section, we note that using a large value of $N$ appears to lead to poor results; it appears likely that the standard deviation of the posterior is being underestimated.

\begin{table}[ht]
\centering
\begin{tabular}{r|rrr|rrr|r}
  \hline
Method & Mean & sd & sd$\times \sqrt{\bar{\mathbf{S}}}$ & Mean & sd & sd$\times \sqrt{\bar{\mathbf{S}}}$ & $\bar{\mathbf{S}}$ \\ 
  \hline
DA $N=500$, $B=1$ & 3.3776 & 0.5517 & 11579 & 2.5534 & 0.4121 & 8649 & $4.41 \times 10^5$ \\ 
DA $N=1000$, $B=1$ & 3.6168 & 1.2165 & 21352 & 2.4516 & 0.7231 & 12693 & $3.08 \times 10^5$ \\ 
DA $N=5000$, $B=1$ & 3.7702 & 1.6169 & 26462 & 1.7376 & 0.9840 & 16104 & $2.68 \times 10^5$ \\ 
DA $N=10000$, $B=1$ & 3.7532 & 2.0874 & 33747 & 1.9016 & 0.8581 & 13874 & $2.61 \times 10^5$ \\ 
DA $N=50000$, $B=1$ & 3.2808 & 2.4644 & 44738 & 1.5644 & 1.0555 & 19162 & $3.30  \times 10^5$ \\ 
DA $N=500$, $B=5$ & 3.8090 & 1.1143 & 22445 & 2.7266 & 0.6132 & 12353 & $4.06 \times 10^5$ \\ 
DA $N=1000$, $B=5$ & 3.4027 & 0.9633 & 16572 & 2.3664 & 0.8668 & 14911 & $2.96 \times 10^5$ \\ 
DA $N=5000$, $B=5$ & 3.5438 & 2.0486 & 35046 & 1.8708 & 0.6549 & 11205 & $2.93 \times 10^5$ \\ 
DA $N=10000$, $B=5$ & 3.6596 & 1.7692 & 31444 & 1.7683 & 0.9247 & 16435 & $3.16 \times 10^5$ \\ 
DA $N=50000$, $B=5$ & 3.8090 & 2.0645 & 43305 & 1.4834 & 1.0414 & 21845 & $4.40 \times 10^5$ \\ 
ABC-SMC $N=200$ & 3.7580 & 0.3659 & 31000 & 2.9472 & 0.2501 & 21188 & $7.18 \times 10^6$ \\ 
   \hline
\end{tabular}
\caption{The mean and standard deviation of the estimated posterior mean (columns 2-4) and standard deviation (columns 5-7) of $\rho$, and the median number $\bar{\mathbf{S}}$ of sweeps of the Gibbs sampler $\bar{\mathbf{S}}$. All rows refer to DA approaches, except the last, which is standard ABC-SMC.} \label{t:rho}
\end{table}






\section{Conclusions\label{sec:Conclusions}}

In this paper we introduced DA-ABC-SMC as a~means for reducing the
computational cost of ABC-SMC in the case of an expensive simulator,
through using a~DA-ABC-MCMC move, in which a~cheap simulator is used
to ``screen'' the candidate values of proposed parameters so that
less effort needs to be used running the expensive simulator. This
cheap simulator may be completely independent of the expensive simulator,
but preferably the expensive simulator may be conditioned on the output
of the cheap simulator. The tolerance for the cheap simulator is chosen
adaptively with DA-ABC-SMC, giving an algorithm that only has three relatively easy to choose tuning parameters. The key factor
in the performance of the algorithm is the DA proposal,
i.e. the distribution of the particles that pass the first stage of
DA. For this proposal to be useful, the cheap simulator needs to produce
a posterior centred around the same values as the expensive
simulator, and the total number of particles $N$ in the DA-ABC-SMC
needs to be chosen appropriately. The empirical results suggest that there is a trade-off in choosing the size of $N$: large values result in the largest computational saving, but can produce a DA proposal that is too concentrated which can result in a poor sample from the posterior. For smaller values of $N$ the DA proposal is usually more suitable, and can result in computational savings compared to ABC-SMC.

Section \ref{sec:Application-to-stochastic} illustrates that DA-ABC-SMC
shows promise for using ABC for inference in ODE or SDE models, where
a numerical method is used to simulate from the model. Section \ref{sec:Application-to-Ising}
revisits the idea of using ABC for inference in latent Markov random
field models, and suggests an approach that in some cases reduces the computational cost of ABC-SMC by using short runs of MCMC to simulate from the model. Future work might extend the approach to ABC methods that do not use an indicator kernel, use the adaptive tuning of the first stage of DA outside of the ABC context (along with an investigation of any bias this adaptation may introduce), or aim to improve the automation of the design of the DA proposal, which is often too concentrated.

\section*{Acknowledgements}

Many thanks to Chris Drovandi for pointing out the method of \citet{Duan2015}, and for discussions on using DA within SMC outside of the ABC setting. Richard Everitt?s work was supported by BBSRC grant BB/N00874X/1 and EPSRC grant EP/N023927/1, and Paulina Rowi{\'n}ska?s work was supported by EPSRC grant EP/L016613/1 (the Centre for Doctoral Training in the Mathematics of Planet Earth).

\appendix

\section{ABC-MCMC and ABC-SMC}

This section derives the ABC-MCMC algorithm (section \ref{sub:abc-mcmc-derive}), and the ABC-SMC of \citet{DelMoral2012g} (section \ref{sub:abc-smc-derive}).

\subsection{ABC-MCMC and early rejection} \label{sub:abc-mcmc-derive}

ABC-MCMC \citet{Marjoram2003} can be derived by applying standard Metropolis-Hastings (MH) to the target

\begin{equation}
\pi \left(\theta,x \mid y\right) \propto p\left(\theta \right)l\left(x \mid\theta \right)P_{\epsilon }\left(y\mid x\right) \label{eq:abc_target}
\end{equation}
(omitting the dependence of the target on $\epsilon$ to keep the notation consistent throughout the paper). ABC-MCMC uses the proposal $q\left( \theta^* \mid \theta \right) l\left( x^* \mid \theta^* \right)$ on the pair $(\theta, x)$, giving the acceptance probability

\begin{eqnarray}
\alpha \left( (\theta, x)  , (\theta^*, x^*) \right) & = & \frac{p\left(\theta^* \right)l\left(x^* \mid \theta^* \right)P_{\epsilon }\left(y\mid x^*\right)}{p\left(\theta \right)l\left(x \mid\theta \right)P_{\epsilon }\left(y\mid x\right)} \frac{q\left( \theta \mid \theta^*\right) l\left( x \mid \theta \right)}{q\left( \theta^* \mid \theta \right) l\left( x^* \mid \theta^* \right)} \\
 & = & \frac{p\left(\theta^* \right) P_{\epsilon }\left(y\mid x^*\right)}{p\left(\theta \right) P_{\epsilon }\left(y\mid x\right)} \frac{q\left( \theta \mid \theta^*\right) }{q\left( \theta^* \mid \theta \right) }.
\end{eqnarray}
The cancellation of the likelihood terms allows this algorithm to be implemented.

\subsubsection{Indicator kernels and early rejection}

In ABC~a common choice for the kernel $P_{\epsilon}$ is $P_{\epsilon}\left(y\mid x\right) \propto \mathbb{I}\left(d\left(y,x\right)<=\epsilon \right)$, where $d$ is~a distance metric. We now consider the implications of using this kernel on the ABC-MCMC acceptance probability in the preceding section. Firstly, we must consider the possibility that the denominator in the ratio is equal to 0, due to having $d\left(y,x\right)>\epsilon$. The general MH framework of \citet{Tierney1998} (see the main paper) dictates that the acceptance probability in this case should be 0: the practical implications of this are that one must ensure that the initial value of $x$ satisfies $d\left(y,x\right)<=\epsilon$, or the chain will never move from its starting point (in practice instead often different values of $(\theta, x)$ are explored until $d\left(y,x\right)<=\epsilon$ is satisfied, with this initial exploration being discarded). Therefore, we may always assume that $d\left(y,x\right)<=\epsilon$ after the chain is initialised and hence the acceptance probability may be written as

\begin{equation}
\alpha \left( (\theta, x)  , (\theta^*, x^*) \right) = \begin{cases}
\frac{p\left(\theta^{*}\right)q\left(\theta\mid\theta^{*}\right)}{p\left(\theta\right)q\left(\theta^{*}\mid\theta\right)} & \text{if }d\left(y,x^{*}\right)<=\epsilon\\
0 & \text{otherwise}
\end{cases}.
\end{equation}
Let $u$ be the uniformly distributed random number in $[0,1]$ generated in order implement the accept-reject step. \citet{Picchini2013} note that a rejection will always occur if

\begin{equation}
u>\frac{p\left(\theta^{*}\right)q\left(\theta\mid\theta^{*}\right)}{p\left(\theta\right)q\left(\theta^{*}\mid\theta\right)} 
\end{equation}
thus this condition may be checked before $x^*$ is simulated from the likelihood. If the proposed point is not rejected after this first step, $x^*$ is simulated, and the proposal is accepted if $d\left(y,x^{*}\right)<=\epsilon$. The consequence of this idea is that for $\theta^*$ that have a small probability under the prior, we have an ``early rejection'' that avoids the (potentially expensive) cost of simulation. The computational savings of this approach will only be significant if the posterior is not too different from the prior.

For clarity, pseudo-code for the early rejection method is given in algorithm \ref{alg:early_rej}.

\begin{algorithm}[H]
\caption{A single iteration of early rejection ABC-MCMC.}
\label{alg:early_rej}
\hspace*{\algorithmicindent} \textbf{Inputs:} Current value of $\theta$. \\
 \hspace*{\algorithmicindent} \textbf{Outputs:} Proposed value $\theta^*$ and accept/reject decision for this proposed value. 
\begin{algorithmic}
\State $u \sim \mathcal{U}\left(0,1\right)$
\State $\theta^* \sim q\left( \cdot \mid \theta \right)$
\If {$u < 1 \wedge \frac{p\left( \theta^* \right) q\left( \theta \mid \theta^* \right) }{ p\left( \theta \right) q\left( \theta^* \mid \theta \right) }$ }
    \State $ x^* \sim l\left( \cdot \mid \theta^* \right)$
    \If { $d\left(y, x^* \right) < \epsilon$ }
        \State Accept $\theta^*$.
    \Else
        \State Reject $\theta^*$.
    \EndIf
\Else
    \State Reject $\theta^*$.
\EndIf
\end{algorithmic}
\end{algorithm}

\subsection{ABC-SMC} \label{sub:abc-smc-derive}

Recall from the main paper the weight update used in an SMC sampler \citep{DelMoral2006c} when using a sequence of targets $\pi_t$, an MCMC move for the ``move'' step, and where the SMC sampler backwards kernel is the reverse of this MCMC kernel:
\begin{eqnarray}
\tilde{w}_{t+1}^{(i)} & = & w_{t}^{(i)}\frac{\pi_{t+1}\left(\theta_{t}^{(i)}\right)}{\pi_{t}\left(\theta_{t}^{(i)}\right)}.\label{eq:smc_weight-1}
\end{eqnarray}
The sequence of targets used in the ABC-SMC sampler of \citet{DelMoral2012g} is given by
\begin{equation}
\pi_{t}\left(\theta_{t},x_{t}\mid y\right) \propto p\left(\theta_{t}\right)l\left(x_{t}\mid\theta_{t}\right)P_{\epsilon_{t}}\left(y\mid x_{t}\right),\label{eq:abc_smc_seq2}
\end{equation}
for $t=1:T$. We saw in the previous section that the ABC-MCMC kernel is a valid MCMC kernel targeting $\pi_{t}\left(\theta_{t},x_{t}\mid y\right)$. Choosing the SMC sampler backwards kernel to be the reverse of this MCMC kernel we obtain
\begin{eqnarray}
\tilde{w}_{t+1} & = & w_{t} \frac{p\left(\theta_{t}\right)l\left(x_{t}\mid\theta_{t}\right)P_{\epsilon_{t+1}}\left(y\mid x_{t}\right)}{p\left(\theta_{t}\right)l\left(x_{t}\mid\theta_{t}\right)P_{\epsilon_{t}}\left(y\mid x_{t}\right)} \\
 & = & w_{t}\frac{P_{\epsilon_{t+1}}\left(y\mid x_{t}\right)}{P_{\epsilon_{t}}\left(y\mid x_{t}\right)}.
\end{eqnarray}

In the case of indicator kernels, this weight update becomes

\[
\tilde{w}_{t+1}=w_{t}\mathbb{I}\left(d\left(y,x_{t}\right)<\epsilon_{t+1}\right),
\]
and early rejection may be used in the ABC-MCMC move. Early rejection may provide a significant computational saving in the early stages of the SMC, since the target is likely to be close to the prior.

\section{Derivation of DA-ABC-MCMC and DA-ABC-SMC}

\subsection{DA-ABC-MCMC}

We present a derivation of DA-ABC-MCMC, using the notation from the main paper. As in the previous sections, the extended state space view of ABC-MCMC is used, where the move is
seen to be a~Metropolis-Hastings move on the space $\left(\theta,x_{1}\right)$,
with $\theta^{*}$ proposed via $q\left(\cdot\mid\theta\right)$
and $x_{1}^{*}$ via $l_{1}\left(\cdot\mid\theta^{*}\right)$. We may view the move as being on the space $\left(\theta,x_{1},x_{2}\right)$, with target proportional to
\[
p\left(\theta^{*}\right) l_{1}\left(x_{1} \mid\theta \right) l_{2}\left(x_{2} \mid x_{1} ,\theta \right) P_{\epsilon_{2}}\left(y\mid x_{1} ,x_{2} \right)
\]
with $x_{2}^{*}$ being proposed via $l_{2}\left(\cdot\mid x_{1}^{*},\theta^{*}\right)$.
We will see that in practice this simulation does not need
to be performed at the first stage (this construction is essentially
the same as in \citet{Sherlock2017}).
The acceptance probability at the first stage is
\begin{eqnarray*}
\alpha_{1} & = & \min\left\{ 1,\frac{p\left(\theta^{*}\right)l_{1}\left(x_{1}^{*}\mid\theta^{*}\right)P_{\epsilon_{1}}\left(y_{1}\mid x_{1}^{*}\right)l_{2}\left(x_{2}^{*}\mid x_{1}^{*},\theta^{*}\right)}{p\left(\theta\right)l_{1}\left(x_{1}\mid\theta\right)P_{\epsilon_{1}}\left(y_{1}\mid x_{1}\right)l_{2}\left(x_{2}\mid x_{1},\theta\right)}\frac{q\left(\theta\mid\theta^{*}\right)l_{1}\left(x_{1}\mid\theta\right)l_{2}\left(x_{2}\mid x_{1},\theta\right)}{q\left(\theta^{*}\mid\theta\right)l_{1}\left(x_{1}^{*}\mid\theta^{*}\right)l_{2}\left(x_{2}^{*}\mid x_{1}^{*},\theta^{*}\right)}\right\} \\
 & = & \min\left\{ 1,\frac{p\left(\theta^{*}\right)P_{\epsilon_{1}}\left(y_{1}\mid x_{1}^{*}\right)}{p\left(\theta\right)P_{\epsilon_{1}}\left(y_{1}\mid x_{1}\right)}\frac{q\left(\theta\mid\theta^{*}\right)}{q\left(\theta^{*}\mid\theta\right)}\right\}.
\end{eqnarray*}
We observe that the marginal distribution
of the target we have used is the ABC posterior with $l_{1}$, $\epsilon_{1}$
and $y_{1}$.

Using delayed acceptance, as described in the main paper, the acceptance
probability at the second stage is
\begin{eqnarray*}
\alpha_{2} & = & \min\left\{ 1,\frac{p\left(\theta^{*}\right)l_{1}\left(x_{1}^{*}\mid\theta^{*}\right)l_{2}\left(x_{2}^{*}\mid x_{1}^{*},\theta^{*}\right)P_{\epsilon_{2}}\left(y\mid x_{1}^{*},x_{2}^{*}\right)}{p\left(\theta\right)l_{1}\left(x_{1}\mid\theta\right)l_{2}\left(x_{2}\mid x_{1},\theta\right)P_{\epsilon_{2}}\left(y\mid x_{1},x_{2}\right)}\frac{p\left(\theta\right)l_{1}\left(x_{1}\mid\theta\right)P_{\epsilon_{1}}\left(y_{1}\mid x_{1}\right)l_{2}\left(x_{2}\mid x_{1},\theta\right)}{p\left(\theta^{*}\right)l_{1}\left(x_{1}^{*}\mid\theta^{*}\right)P_{\epsilon_{1}}\left(y_{1}\mid x_{1}^{*}\right)l_{2}\left(x_{2}^{*}\mid x_{1}^{*},\theta^{*}\right)}\right\} \\
 & = & \min\left\{ 1,\frac{P_{\epsilon_{2}}\left(y\mid x_{1}^{*},x_{2}^{*}\right)}{P_{\epsilon_{2}}\left(y\mid x_{1},x_{2}\right)}\frac{P_{\epsilon_{1}}\left(y_{1}\mid x_{1}\right)}{P_{\epsilon_{1}}\left(y_{1}\mid x_{1}^{*}\right)}\right\}.
\end{eqnarray*}

\subsection{DA-ABC-SMC}

We now justify the weight update for the DA-ABC-SMC method described in the main paper. The target distribution used at iteration $t$ is proportional to
\[
p\left(\theta_{t}\right)l_{1}\left(x_{1,t}\mid\theta_{t}\right)l_{2}\left(x_{2,t}\mid x_{1,t},\theta_{t}\right)P_{\epsilon_{2,,t}}\left(y\mid x_{1,t},x_{2,t}\right).
\]
Using the same approach as in section \ref{sub:abc-smc-derive}, we see that the weight update for each particle is given by
\begin{eqnarray*}
\tilde{w}_{t+1} & = & w_{t}\frac{p\left(\theta_{t}\right)l_{1}\left(x_{1,t}\mid\theta_{t}\right)l_{2}\left(x_{2,t}\mid x_{1,t},\theta_{t}\right)P_{\epsilon_{2,,t+1}}\left(y\mid x_{1,t},x_{2,t}\right)}{p\left(\theta_{t}\right)l_{1}\left(x_{1,t}\mid\theta_{t}\right)l_{2}\left(x_{2,t}\mid x_{1,t},\theta_{t}\right)P_{\epsilon_{2,,t}}\left(y\mid x_{1,t},x_{2,t}\right)}\\
 & = & w_{t}\frac{P_{\epsilon_{2,,t+1}}\left(y\mid x_{1,t},x_{2,t}\right)}{P_{\epsilon_{2,,t}}\left(y\mid x_{1,t},x_{2,t}\right)}.
\end{eqnarray*}

\subsubsection{Using indicator kernels}

In this section we consider the situation when $P_{\epsilon_{1,t}}$
is chosen to be an indicator function; i.e. $P_{\epsilon_{1,t}}\left(y_{1,t}\mid x_{1,t}^{*}\right)\propto\mathbb{I}\left(d\left(y_{1,t},x_{1,t}^{*}\right)<=\epsilon_{1,t}\right)$,
where $d$ is a~distance metric. In this case when specifying our
acceptance probabilities we need to account for our target distributions
having zero density in some parts of the space. We follow the framework
of \citet{Tierney1998}, in which for a~target $\pi(\theta)$ and
proposal $q$, the MH acceptance probability is written as
\[
\alpha=\begin{cases}
\min\left\{ 1,\frac{\pi\left(\theta^{*}\right)q\left(\theta\mid\theta^{*}\right)}{\pi\left(\theta\right)q\left(\theta^{*}\mid\theta\right)}\right\}  & \left(\theta,\theta^{*}\right)\in R,\\
0 & \left(\theta,\theta^{*}\right)\notin R,
\end{cases}
\]
where $R=\left\{ \left(\theta,\theta^{*}\right)\mid\pi\left(\theta^{*}\right)q\left(\theta\mid\theta^{*}\right)>0,\pi\left(\theta\right)q\left(\theta^{*}\mid\theta\right)>0\right\} $.
Thus, at the $t$-th iteration of the SMC the acceptance probability
at the first stage of the delayed acceptance is
\begin{equation}
\alpha_{1,t}=\begin{cases}
\min\left\{ 1,\frac{p\left(\theta_{t}^{*}\right)}{p\left(\theta_{t}\right)}\frac{q\left(\theta_{t}\mid\theta_{t}^{*}\right)}{q\left(\theta_{t}^{*}\mid\theta_{t}\right)}\right\}  & d\left(y_{1,t},x_{1,t}\right),d\left(y_{1,t},x_{1,t}^{*}\right)<\epsilon_{1,t},\\
0 & \mbox{otherwise.}
\end{cases}.\label{eq:1st_stage_acc_smc}
\end{equation}
As in \citet{Picchini2013}, we may perform the first stage of delayed
acceptance in two stages, which we will refer to as steps 1a and 1b,
such that some simulations from $l_{1}$ may be avoided. At step 1a,
$\theta_{t}^{*}$ is simulated from $q\left(\cdot\mid\theta_{t}\right)$
and an accept-reject step is performed using the acceptance probability
\[
\alpha_{1\mbox{a},t}=\min\left\{ 1,\frac{p\left(\theta_{t}^{*}\right)}{p\left(\theta_{t}\right)}\frac{q\left(\theta_{t}\mid\theta_{t}^{*}\right)}{q\left(\theta_{t}^{*}\mid\theta_{t}\right)}\right\}.
\]
At step 1b, $x_{1,t}^{*}$ is simulated from $l_{1}\left(\cdot\mid\theta_{t}^{*}\right)$
and the entire move $\left(\theta_{t}^{*},x_{1,t}^{*}\right)$ is
accepted (to be used in stage 2) with probability
\begin{equation}
\alpha_{1b,t}=\begin{cases}
1 & d\left(y_{1,t},x_{1,t}\right),d\left(y_{1,t},x_{1,t}^{*}\right)<\epsilon_{1,t}\\
0 & \mbox{otherwise}
\end{cases}.\label{eq:1bth_stage_acc_smc}
\end{equation}
Splitting the first stage into two substages could itself be seen
as a~form of delayed acceptance, but its acceptance rate is the same
as the single step implementation since it simply uses the fact that
$\mathbb{I}\left(d\left(y_{1,t},x_{1,t}^{*}\right)<\epsilon_{1,t}\right)$
is either 1 or 0 in order to reorganise the single step calculation
in a~computationally efficient way.

The acceptance probability at the second stage is
\[
\alpha_{2,t}=\begin{cases}
1 & d\left(y_{1,t},x_{1,t}\right),d\left(y_{1,t},x_{1,t}^{*}\right)<\epsilon_{1,t}\mbox{ and }d\left(y,x_{2,t}\right),d\left(y,x_{2,t}^{*}\right)<\epsilon_{2,t}\\
0 & \mbox{otherwise}
\end{cases},
\]
which may be seen directly from the description of DA from the main paper
with the appropriate choices of $\pi_{2}$ and $K_{1}$. Note
that $d\left(y,x_{2,t}\right)<\epsilon_{2,t}$ must be true for the
particle to have non-zero weight, and $d\left(y_{1,t},x_{1,t}\right),d\left(y_{1,t},x_{1,t}^{*}\right)<\epsilon_{1,t}$
must be satisfied to reach the second stage of DA, thus in practice
we use
\[
\alpha_{2,t}=\begin{cases}
1 & d\left(y,x_{2,t}^{*}\right)<\epsilon_{2,t},\\
0 & \mbox{otherwise.}
\end{cases}
\]



\section{Lotka-Volterra model}

\subsection{Full details of methods}

All of our empirical results were generated using R \citep{R2019}. We study the data ``\texttt{LVPerfect}'' in the R package \texttt{smfsb} \citep{smfsb} (the numerical
methods for simulating from the likelihood are also taken from this
package), previously studied in \citet{Wilkinson2011}. In this data
the simulation starts with initial populations $X=50$ and $Y=100$,
and has 30 time units, with the values of $X$ and $Y$ being recorded
every 2 time units, resulting in 16 data points in each of the two
time series. Our prior follows that in \citet{Wilkinson2011}, being
uniform in the $\log$ domain. Specifically we use
\[
p\left(\log\left(\theta\right)\right)\propto\prod_{i=1}^{3}\mathcal{U}\left(\log\left(\theta_{i}\right)\mid\text{lower}=-6,\text{upper}=2\right).
\]

Our ABC approach follows that in \citet{Wilkinson2011,Papamakarios2016}:
as summary statistics we use a~9-dimensional vector composed of the
mean, $\log$ variance and first two autocorrelations of each time
series, together with the cross-correlation between them. These statistics
were normalised by the standard deviation of the statistics determined
by a~pilot run, precisely as in \citet{Wilkinson2011}. The distance
between the summary statistics used in ABC was taken to be the Euclidean
norm between the normalised statistic vectors. In all our ABC algorithms
we used a~final tolerance $\log (\epsilon_{2})= \log(0.15) \approx -1.89712$. Reducing the tolerance
below this level does not appear to have a~large impact on the posterior
distribution.

We used DA-ABC-SMC with a variety of choices of $U$, $A$ and $N$,
and two different choices of the Euler-Maruyama step size $s$ in
the cheap simulator $s=0.5$ and $s=0.1$, both of which result in very rough approximations of the dynamics.  We compared these approaches with standard ABC-SMC,
with $N=200$ particles and a~sequence of tolerances selected
such that $U=100$ unique particles are retained at each iteration, and also ``ground truth'' for the posterior expectation and standard deviation of the parameters found using a long run ($10^5$ iterations) of ABC-MCMC. We also compared our approach with a method based on the SMC$^2$-style approach of \citet{Duan2015}. This approach uses the same SMC-based likelihood estimate (with $M$ particles) as particle MCMC \citep{Wilkinson2011}, but embeds this within an SMC sampler rather than a pseudo-marginal MCMC chain. The sequence of distributions in the ``external'' SMC sampler is given by raising the likelihood estimate to a power: beginning with 0 and ending with 1 (so that the final distribution is the true posterior). The posterior targeted by this method is the same as in particle MCMC. We used the same model as in the particle MCMC of \citet{Wilkinson2011}: specifically we used a normal distribution with mean 0 and standard deviation 10 as the measurement model at each time step. \citet{Wilkinson2011} shows that the posterior obtained using this model has a much smaller standard deviation compared to the one obtained when using ABC, therefore when comparing the SMC$^2$ approach with ABC, we only compared the posterior mean (bearing in mind that this is also slightly different between the two cases). In order that the computational cost is comparable with the ABC approaches, we use a DA move within the method of \citet{Duan2015}. Full details of this method follow.

Algorithm \ref{a:duan} gives the SMC sampler of \citet{Duan2015}. This method is adapted so that the sequence of powers to which the likelihood estimates are raised is determined adaptively, by using a bisection search to find the power such that the conditional effective sample size (CESS) \citep{Zhou2015} is $\alpha N$ (where $\alpha$ is a proportion). The CESS is defined as
\[
\text{CESS}=\frac{N\left(\sum_{i=1}^{N}w_{t+1}^{(i)}\omega_t^{(i)}\right)^{2}}{\sum_{i=1}^{N}w_{t+1}^{(i)}\left(\omega_t^{(i)}\right)^{2}}
\]
where $\omega^{(i)}$ is the incremental weight for the $i$th particle (the factor by which we multiply $w_{t}^{(i)}$ by in order to obtain $\tilde{w}_{t+1}^{(i)}$.

\begin{algorithm}[H]
\caption{The SMC sampler of \citet{Duan2015}, with adaptation to choose the sequence of distributions.}
\label{a:duan}
\hspace*{\algorithmicindent} \textbf{Inputs:} Number of particles $N$, the proportion $\alpha$ used in the adaptive approach to choosing the sequence of distributions, the proportion $\beta$ used in resampling, prior $p$, particle filtering parameters for estimating $l$ (including the number of particles $M$). \\
 \hspace*{\algorithmicindent} \textbf{Outputs:} Particles $\left\{\left( \theta^{(i)}_t,x^{(i)}_t \right) \right\}_{i=1}^N$ and weights $\left\{w_{t}^{(i)} \right\}_{i=1}^N$ for all $t$.
\begin{algorithmic}
\For {$i= 1:N$}
    \State $\theta^{(i)}_0 \sim p\left( \cdot \right)$
    \State Run a particle filter to find the likelihood estimate $\hat{l}^{(i)}_0$ at $\theta^{(i)}_0$.
    \State $w^{(i)}_0 = 1/N$
\EndFor
\State $\tau_0=0$, $t=0$.
\While {$\tau_t < 1$}
    \State Use bisection to choose $\tau_{t+1}$ s.t. the CESS is $\alpha N$.
    \For {$i= 1:N$}
        \State $\tilde{w}_{t+1}^{(i)}=w_{t}^{(i)} (\hat{l}^{(i)}_t)^{(\tau_{t+1}-\tau_t)}$
    \EndFor
    \State Normalise $\left\{ \tilde{w}_{t+1} \right\}_{i=1}^N$ to give normalised weights $\left\{ w_{t+1} \right\}_{i=1}^N$.
    \State Perform resampling if the ESS falls below $\beta N$.
    \For {$i= 1:N$}
        \State $\theta^{(i)}_{t+1} =  \theta^{(i)}_{t}$, $\hat{l}^{(i)}_{t+1} =  \hat{l}^{(i)}_{t}$
        \State $\left(\theta^{(i)}_{t+1}\right)^* \sim q\left( \cdot \mid \theta^{(i)}_{t} \right)$
        \State Run a particle filter to find the likelihood estimate $\left( \hat{l}_{t+1}^{(i)} \right)^*$ at $\left(\theta^{(i)}_{t+1}\right)^*$.
        \State $u \sim \mathcal{U}\left(0,1\right)$
        \If {$u < 1 \wedge \frac{p\left( \left(\theta^{(i)}_{t+1}\right)^* \right) \left( \left(\hat{l}^{(i)}_{t+1}\right)^* \right)^{\tau_{t+1}} q\left( \theta^{(i)}_{t} \mid \left(\theta^{(i)}_{t+1}\right)^* \right) }{ p\left( \theta^{(i)}_{t+1} \right) \left( \hat{l}^{(i)}_{t+1} \right)^{\tau_{t+1}} q\left( \left(\theta^{(i)}_{t+1}\right)^* \mid \theta^{(i)}_{t} \right) }$}
            \State $\theta^{(i)}_{t+1} =  \left(\theta^{(i)}_{t+1}\right)^*$, $\hat{l}^{(i)}_{t+1} =  \left(\hat{l}^{(i)}_{t+1}\right)^*$
        \EndIf
    \EndFor
    \State $t = t + 1$
\EndWhile
\end{algorithmic}
\end{algorithm}

When applying algorithm \ref{a:duan} to the Lotka-Volterra data, we found that in order for the SMC to avoid degeneracy (and give a posterior near to the true posterior), it required a configuration (in terms of choosing appropriate $N$, $\alpha$ and $\beta$) that resulted in a computational cost of more than an order of magnitude slower than the ABC approaches. Due to this, we used a delayed acceptance MCMC move in place of the particle MCMC move given in algorithm \ref{a:duan}. Algorithm \ref{a:duan_da} gives the resultant algorithm. In this approach, analogous to our description of DA-ABC-SMC, $l_2=l$ is the true  likelihood to be estimated using a particle filter, and $l_1$ is a computationally cheap likelihood. Algorithm \ref{a:duan_da} was the method used in the main paper, with $\beta=0.5$ in all cases, and different values of $N$, $M$ and $\alpha$. We note that the SMC$^2$ method of \citet{Chopin2013} was also tried, but was not found to be competitive in terms of the computational effort required to avoid degeneracy.

\begin{algorithm}[H]
\caption{The SMC sampler of \citet{Duan2015}, with adaptation to choose the sequence of distributions and a DA-MCMC move.}
\label{a:duan_da}
\hspace*{\algorithmicindent} \textbf{Inputs:} Number of particles $N$, the proportion $\alpha$ used in the adaptive approach to choosing the sequence of distributions, the proportion $\beta$ used in resampling, prior $p$, particle filtering parameters for estimating $l_1$ and $l_2=l$ (including the number of particles $M$). \\
 \hspace*{\algorithmicindent} \textbf{Outputs:} Particles $\left\{\left( \theta^{(i)}_t,x^{(i)}_t \right) \right\}_{i=1}^N$ and weights $\left\{w_{t}^{(i)} \right\}_{i=1}^N$ for all $t$.
\begin{algorithmic}
\For {$i= 1:N$}
    \State $\theta^{(i)}_0 \sim p\left( \cdot \right)$
    \State Run particle filters to find the likelihood estimates $\hat{l}^{(i)}_{1,0}$ and $\hat{l}^{(i)}_{2,0}$ at $\theta^{(i)}_0$.
    \State $w^{(i)}_0 = 1/N$
\EndFor
\State $\tau_0=0$, $t=0$.
\While {$\tau_t < 1$}
    \State Use bisection to choose $\tau_{t+1}$ s.t. the CESS is $\alpha N$.
    \For {$i= 1:N$}
        \State $\tilde{w}_{t+1}^{(i)}=w_{t}^{(i)} (\hat{l}^{(i)}_{2,t})^{(\tau_{t+1}-\tau_t)}$
    \EndFor
    \State Normalise $\left\{ \tilde{w}_{t+1} \right\}_{i=1}^N$ to give normalised weights $\left\{ w_{t+1} \right\}_{i=1}^N$.
    \State Perform resampling if the ESS falls below $\beta N$.
    \For {$i= 1:N$}
        \State $\theta^{(i)}_{t+1} =  \theta^{(i)}_{t}$, $\hat{l}^{(i)}_{t+1} =  \hat{l}^{(i)}_{1,t}$
        \State $\left(\theta^{(i)}_{t+1}\right)^* \sim q\left( \cdot \mid \theta^{(i)}_{t} \right)$
        \State Run a particle filter to find the likelihood estimate $\left( \hat{l_1}_{t+1}^{(i)} \right)^*$ at $\left(\theta^{(i)}_{t+1}\right)^*$.
        \State $u_1 \sim \mathcal{U}\left(0,1\right)$
        \If {$u_1 < 1 \wedge \frac{p\left( \left(\theta^{(i)}_{t+1}\right)^* \right) \left( \left( \hat{l}^{(i)}_{1,t+1}\right)^* \right)^{\tau_{t+1}} q\left( \theta^{(i)}_{t} \mid \left(\theta^{(i)}_{t+1}\right)^* \right) }{ p\left( \theta^{(i)}_{t+1} \right) \left( \hat{l}^{(i)}_{1,t+1} \right)^{\tau_{t+1}} q\left( \left(\theta^{(i)}_{t+1}\right)^* \mid \theta^{(i)}_{t} \right) }$}
            \State Run a particle filter to find the likelihood estimate $\left( \hat{l_2}_{t+1}^{(i)} \right)^*$ at $\left(\theta^{(i)}_{t+1}\right)^*$.
             \State $u_2 \sim \mathcal{U}\left(0,1\right)$
             \If {$u_2 < 1 \wedge \frac{ \left( \left( \hat{l}^{(i)}_{2,t+1}\right)^* \right)^{\tau_{t+1}} \left( \hat{l}^{(i)}_{1,t+1} \right)^{\tau_{t+1}} }{ \left( \hat{l}^{(i)}_{2,t+1} \right)^{\tau_{t+1}} \left( \left( \hat{l}^{(i)}_{1,t+1}\right)^* \right)^{\tau_{t+1}} }$  }
                \State $\theta^{(i)}_{t+1} =  \left(\theta^{(i)}_{t+1}\right)^*$, $\hat{l}^{(i)}_{1,t+1} =  \left(\hat{l}^{(i)}_{1,t+1}\right)^*$, $\hat{l}^{(i)}_{2,t+1} =  \left(\hat{l}^{(i)}_{2,t+1}\right)^*$
            \EndIf
        \EndIf
    \EndFor
    \State $t = t + 1$
\EndWhile
\end{algorithmic}
\end{algorithm}

All algorithms were run 30 times, and used an expensive simulator with Euler-Maruyama step
size $0.0005$ (which
resulted in a~very accurate approximation), and included the scheme of \citet{Picchini2013} to
avoid simulations from the likelihood where they may be rejected using
the prior only. In all approaches the MCMC proposal was Gaussian
centred at the current point with variance given by the sample variance
of the previous particles. To measure computational cost, we counted
the total number of steps $\mathbf{S}$ (taking the median $\bar{\mathbf{S}}$ over the 30 runs) simulated using Euler-Maruyama, taking into account that some simulations were cut short due to the numerical solver diverging (in the implementation in the \texttt{smfsb} package, the practical result of this is that after a certain point in the simulation, the size of the populations is assigned ``NaN''). When the simulation diverged, both population sizes were assigned to be zero after the time of the divergence.

\subsection{Results}

The R packages \texttt{ggplot2} \citep{ggplot2}, \texttt{matlab} \citep{matlab} and \texttt{mvtnorm} \citep{mvtnorm} were used when generating the results for this section. Our first observation is with the parameters $N=200$ and $U=100$, ABC-SMC sometimes had difficulty converging to the final tolerance. Of the 30 runs, 4 runs were not close to reducing the log tolerance to $0.15$ (for some runs this was the case after more than 20,000 SMC iterations). This was not observed for any other approach. In order to present comparisons between ABC-SMC and the other approaches, we focus on median rather than mean simulation times (reported in table \ref{t:times}). We truncated the unfinished runs to 5,000 SMC iterations and treat them as if they had finished, but to provide a fair comparison we also present results where these runs are excluded.

\begin{table}[ht]
\centering
\begin{tabular}{rrr}
  \hline
Method & $\bar{\mathbf{S}}$ & Med. SMC iter. \\ 
  \hline
  DA-ABC-SMC: $N=500$, $U=100$, $A=100$, $s=0.1$.& $1.27 \times 10^9$ & 508 \\ 
  DA-ABC-SMC: $N=500$, $U=100$, $A=100$, $s=0.5$. & $1.08 \times 10^9$ & 231 \\ 
  DA-ABC-SMC: $N=1000$, $U=100$, $A=100$, $s=0.1$. & $8.27 \times 10^8$ & 159.5 \\ 
  DA-ABC-SMC: $N=1000$, $U=100$, $A=100$, $s=0.5$. & $7.04 \times 10^8$ & 159 \\ 
  DA-ABC-SMC: $N=1000$, $U=200$, $A=100$, $s=0.1$. & $5.98 \times 10^9$ & 1423 \\ 
  DA-ABC-SMC: $N=1000$, $U=50$, $A=100$, $s=0.1$. & $2.66 \times 10^8$ & 43.5 \\ 
  DA-ABC-SMC: $N=5000$, $U=100$, $A=100$, $s=0.1$. & $4.78 \times 10^8$ & 65.5 \\ 
  DA-ABC-SMC: $N=10000$, $U=100$, $A=100$, $s=0.1$. & $5.46 \times 10^8$ & 63 \\ 
  SMC$^2$: $N=100$, $M=100$, $\alpha=0.9$, $s=0.1$. & $7.89 \times 10^8$ & 20 \\ 
  SMC$^2$: $N=100$, $M=100$, $\alpha=0.99$, $s=0.1$. & $2.53 \times 10^9$ & 62.5 \\ 
  SMC$^2$: $N=100$, $M=1000$, $\alpha=0.9$, $s=0.1$. & $6.23 \times 10^9$ & 17 \\ 
  ABC-SMC: $N=200$, $U=100$. & $2.84 \times 10^9$ & 816.5 \\ 
   \hline
\end{tabular}
\caption{Median number $\bar{\mathbf{S}}$ of Euler-Maruyama steps and median number of SMC iterations for each method.} \label{t:times}
\end{table}


\begin{figure}[H]
\centering
\begin{subfigure}{.3\textwidth}
  \centering
  \includegraphics[width=1\linewidth]{{{da_epsilon_5e-04_0.1_500_100_0.2_0.15_}}}
  \caption{DA-ABC-SMC: $N=500$, $U=100$, $A=100$, $s=0.1$.}
\end{subfigure}%
\hspace{0.5cm}
\begin{subfigure}{.3\textwidth}
  \centering
  \includegraphics[width=1\linewidth]{{{da_epsilon_5e-04_0.5_500_100_0.2_0.15_}}}
  \caption{DA-ABC-SMC: $N=500$, $U=100$, $A=100$, $s=0.5$.}
\end{subfigure}%
\hspace{0.5cm}
\begin{subfigure}{.3\textwidth}
  \centering
  \includegraphics[width=1\linewidth]{{{da_epsilon_5e-04_0.1_1000_100_0.1_0.15_}}}
  \caption{DA-ABC-SMC: $N=1000$, $U=100$, $A=100$, $s=0.1$.}
\end{subfigure}
\begin{subfigure}{.3\textwidth}
  \centering
  \includegraphics[width=1\linewidth]{{{da_epsilon_5e-04_0.5_1000_100_0.1_0.15_}}}
  \caption{DA-ABC-SMC: $N=1000$, $U=100$, $A=100$, $s=0.5$.}
\end{subfigure}%
\hspace{0.5cm}
\begin{subfigure}{.3\textwidth}
  \centering
  \includegraphics[width=1\linewidth]{{{da_epsilon_5e-04_0.1_1000_100_0.2_0.15_}}}
  \caption{DA-ABC-SMC: $N=1000$, $U=200$, $A=100$, $s=0.1$.}
\end{subfigure}%
\hspace{0.5cm}
\begin{subfigure}{.3\textwidth}
  \centering
  \includegraphics[width=1\linewidth]{{{da_epsilon_5e-04_0.1_1000_100_0.05_0.15_}}}
  \caption{DA-ABC-SMC: $N=1000$, $U=50$, $A=100$, $s=0.1$.}
\end{subfigure}
\begin{subfigure}{.3\textwidth}
  \centering
  \includegraphics[width=1\linewidth]{{{da_epsilon_5e-04_0.1_5000_100_0.02_0.15_}}}
  \caption{DA-ABC-SMC: $N=5000$, $U=100$, $A=100$, $s=0.1$.}
\end{subfigure}%
\hspace{0.5cm}
\begin{subfigure}{.3\textwidth}
  \centering
  \includegraphics[width=1\linewidth]{{{da_epsilon_5e-04_0.1_10000_100_0.01_0.15_}}}
  \caption{DA-ABC-SMC: $N=10000$, $U=100$, $A=100$, $s=0.1$.}
\end{subfigure}
\caption{$\log(\epsilon_1)$, the tolerance in the first stage of DA, plotted against $\log(\epsilon_2)$, the tolerance in the second stage of DA.}
\end{figure}

\begin{figure}[H]
\centering
\begin{subfigure}{.3\textwidth}
  \centering
  \includegraphics[width=1\linewidth]{{{mean1_v_sims_5e-04_0.1_500_100_0.2_0.15_}}}
  \caption{DA-ABC-SMC: $N=500$, $U=100$, $A=100$, $s=0.1$.}
\end{subfigure}%
\hspace{0.5cm}
\begin{subfigure}{.3\textwidth}
  \centering
  \includegraphics[width=1\linewidth]{{{mean1_v_sims_5e-04_0.5_500_100_0.2_0.15_}}}
  \caption{DA-ABC-SMC: $N=500$, $U=100$, $A=100$, $s=0.5$.}
\end{subfigure}%
\hspace{0.5cm}
\begin{subfigure}{.3\textwidth}
  \centering
  \includegraphics[width=1\linewidth]{{{mean1_v_sims_5e-04_0.1_1000_100_0.1_0.15_}}}
  \caption{DA-ABC-SMC: $N=1000$, $U=100$, $A=100$, $s=0.1$.}
\end{subfigure}
\begin{subfigure}{.3\textwidth}
  \centering
  \includegraphics[width=1\linewidth]{{{mean1_v_sims_5e-04_0.5_1000_100_0.1_0.15_}}}
  \caption{DA-ABC-SMC: $N=1000$, $U=100$, $A=100$, $s=0.5$.}
\end{subfigure}%
\hspace{0.5cm}
\begin{subfigure}{.3\textwidth}
  \centering
  \includegraphics[width=1\linewidth]{{{mean1_v_sims_5e-04_0.1_1000_100_0.2_0.15_}}}
  \caption{DA-ABC-SMC: $N=1000$, $U=200$, $A=100$, $s=0.1$.}
\end{subfigure}%
\hspace{0.5cm}
\begin{subfigure}{.3\textwidth}
  \centering
  \includegraphics[width=1\linewidth]{{{mean1_v_sims_5e-04_0.1_1000_100_0.05_0.15_}}}
  \caption{DA-ABC-SMC: $N=1000$, $U=50$, $A=100$, $s=0.1$.}
\end{subfigure}
\begin{subfigure}{.3\textwidth}
  \centering
  \includegraphics[width=1\linewidth]{{{mean1_v_sims_5e-04_0.1_5000_100_0.02_0.15_}}}
  \caption{DA-ABC-SMC: $N=5000$, $U=100$, $A=100$, $s=0.1$.}
\end{subfigure}%
\hspace{0.5cm}
\begin{subfigure}{.3\textwidth}
  \centering
  \includegraphics[width=1\linewidth]{{{mean1_v_sims_5e-04_0.1_10000_100_0.01_0.15_}}}
  \caption{DA-ABC-SMC: $N=10000$, $U=100$, $A=100$, $s=0.1$.}
\end{subfigure}%
\hspace{0.5cm}
\begin{subfigure}{.3\textwidth}
  \centering
  \includegraphics[width=1\linewidth]{{{mean1_v_sims_5e-04_0.1_100_100_0.9_0.5_}}}
  \caption{SMC$^2$: $N=100$, $M=100$, $\alpha=0.9$, $s=0.1$.}
\end{subfigure}
\begin{subfigure}{.3\textwidth}
  \centering
  \includegraphics[width=1\linewidth]{{{mean1_v_sims_5e-04_0.1_100_100_0.99_0.5_}}}
  \caption{SMC$^2$: $N=100$, $M=100$, $\alpha=0.99$, $s=0.1$.}
\end{subfigure}%
\hspace{0.5cm}
\begin{subfigure}{.3\textwidth}
  \centering
  \includegraphics[width=1\linewidth]{{{mean1_v_sims_5e-04_0.1_100_1000_0.9_0.5_}}}
  \caption{SMC$^2$: $N=100$, $M=1000$, $\alpha=0.9$, $s=0.1$.}
\end{subfigure}%
\hspace{0.5cm}
\begin{subfigure}{.3\textwidth}
  \centering
  \includegraphics[width=1\linewidth]{{{mean1_v_sims_5e-04_0.1_200_100_0.5_0.15_}}}
  \caption{ABC-SMC: $N=200$, $U=100$.}
\end{subfigure}
\caption{The estimated posterior mean of $\theta_1$ plotted against the total number of time steps used in Euler-Maruyama. Ground truth from ABC-MCMC is marked with a horizontal line.}
\end{figure}

\begin{figure}[H]
\centering
\begin{subfigure}{.3\textwidth}
  \centering
  \includegraphics[width=1\linewidth]{{{sd1_v_sims_5e-04_0.1_500_100_0.2_0.15_}}}
  \caption{DA-ABC-SMC: $N=500$, $U=100$, $A=100$, $s=0.1$.}
\end{subfigure}%
\hspace{0.5cm}
\begin{subfigure}{.3\textwidth}
  \centering
  \includegraphics[width=1\linewidth]{{{sd1_v_sims_5e-04_0.5_500_100_0.2_0.15_}}}
  \caption{DA-ABC-SMC: $N=500$, $U=100$, $A=100$, $s=0.5$.}
\end{subfigure}%
\hspace{0.5cm}
\begin{subfigure}{.3\textwidth}
  \centering
  \includegraphics[width=1\linewidth]{{{sd1_v_sims_5e-04_0.1_1000_100_0.1_0.15_}}}
  \caption{DA-ABC-SMC: $N=1000$, $U=100$, $A=100$, $s=0.1$.}
\end{subfigure}
\begin{subfigure}{.3\textwidth}
  \centering
  \includegraphics[width=1\linewidth]{{{sd1_v_sims_5e-04_0.5_1000_100_0.1_0.15_}}}
  \caption{DA-ABC-SMC: $N=1000$, $U=100$, $A=100$, $s=0.5$.}
\end{subfigure}%
\hspace{0.5cm}
\begin{subfigure}{.3\textwidth}
  \centering
  \includegraphics[width=1\linewidth]{{{sd1_v_sims_5e-04_0.1_1000_100_0.2_0.15_}}}
  \caption{DA-ABC-SMC: $N=1000$, $U=200$, $A=100$, $s=0.1$.}
\end{subfigure}%
\hspace{0.5cm}
\begin{subfigure}{.3\textwidth}
  \centering
  \includegraphics[width=1\linewidth]{{{sd1_v_sims_5e-04_0.1_1000_100_0.05_0.15_}}}
  \caption{DA-ABC-SMC: $N=1000$, $U=50$, $A=100$, $s=0.1$.}
\end{subfigure}
\begin{subfigure}{.3\textwidth}
  \centering
  \includegraphics[width=1\linewidth]{{{sd1_v_sims_5e-04_0.1_5000_100_0.02_0.15_}}}
  \caption{DA-ABC-SMC: $N=5000$, $U=100$, $A=100$, $s=0.1$.}
\end{subfigure}%
\hspace{0.5cm}
\begin{subfigure}{.3\textwidth}
  \centering
  \includegraphics[width=1\linewidth]{{{sd1_v_sims_5e-04_0.1_10000_100_0.01_0.15_}}}
  \caption{DA-ABC-SMC: $N=10000$, $U=100$, $A=100$, $s=0.1$.}
\end{subfigure}%
\hspace{0.5cm}
\begin{subfigure}{.3\textwidth}
  \centering
  \includegraphics[width=1\linewidth]{{{sd1_v_sims_5e-04_0.1_200_100_0.5_0.15_}}}
  \caption{ABC-SMC: $N=200$, $U=100$.}
\end{subfigure}
\caption{The estimated posterior standard deviation of $\theta_1$ plotted against the total number of time steps used in Euler-Maruyama. Ground truth from ABC-MCMC is marked with a horizontal line.}
\end{figure}

\begin{figure}[H]
\centering
\begin{subfigure}{.3\textwidth}
  \centering
  \includegraphics[width=1\linewidth]{{{mean2_v_sims_5e-04_0.1_500_100_0.2_0.15_}}}
  \caption{DA-ABC-SMC: $N=500$, $U=100$, $A=100$, $s=0.1$.}
\end{subfigure}%
\hspace{0.5cm}
\begin{subfigure}{.3\textwidth}
  \centering
  \includegraphics[width=1\linewidth]{{{mean2_v_sims_5e-04_0.5_500_100_0.2_0.15_}}}
  \caption{DA-ABC-SMC: $N=500$, $U=100$, $A=100$, $s=0.5$.}
\end{subfigure}%
\hspace{0.5cm}
\begin{subfigure}{.3\textwidth}
  \centering
  \includegraphics[width=1\linewidth]{{{mean2_v_sims_5e-04_0.1_1000_100_0.1_0.15_}}}
  \caption{DA-ABC-SMC: $N=1000$, $U=100$, $A=100$, $s=0.1$.}
\end{subfigure}
\begin{subfigure}{.3\textwidth}
  \centering
  \includegraphics[width=1\linewidth]{{{mean2_v_sims_5e-04_0.5_1000_100_0.1_0.15_}}}
  \caption{DA-ABC-SMC: $N=1000$, $U=100$, $A=100$, $s=0.5$.}
\end{subfigure}%
\hspace{0.5cm}
\begin{subfigure}{.3\textwidth}
  \centering
  \includegraphics[width=1\linewidth]{{{mean2_v_sims_5e-04_0.1_1000_100_0.2_0.15_}}}
  \caption{DA-ABC-SMC: $N=1000$, $U=200$, $A=100$, $s=0.1$.}
\end{subfigure}%
\hspace{0.5cm}
\begin{subfigure}{.3\textwidth}
  \centering
  \includegraphics[width=1\linewidth]{{{mean2_v_sims_5e-04_0.1_1000_100_0.05_0.15_}}}
  \caption{DA-ABC-SMC: $N=1000$, $U=50$, $A=100$, $s=0.1$.}
\end{subfigure}
\begin{subfigure}{.3\textwidth}
  \centering
  \includegraphics[width=1\linewidth]{{{mean2_v_sims_5e-04_0.1_5000_100_0.02_0.15_}}}
  \caption{DA-ABC-SMC: $N=5000$, $U=100$, $A=100$, $s=0.1$.}
\end{subfigure}%
\hspace{0.5cm}
\begin{subfigure}{.3\textwidth}
  \centering
  \includegraphics[width=1\linewidth]{{{mean2_v_sims_5e-04_0.1_10000_100_0.01_0.15_}}}
  \caption{DA-ABC-SMC: $N=10000$, $U=100$, $A=100$, $s=0.1$.}
\end{subfigure}%
\hspace{0.5cm}
\begin{subfigure}{.3\textwidth}
  \centering
  \includegraphics[width=1\linewidth]{{{mean2_v_sims_5e-04_0.1_100_100_0.9_0.5_}}}
  \caption{SMC$^2$: $N=100$, $M=100$, $\alpha=0.9$, $s=0.1$.}
\end{subfigure}
\begin{subfigure}{.3\textwidth}
  \centering
  \includegraphics[width=1\linewidth]{{{mean2_v_sims_5e-04_0.1_100_100_0.99_0.5_}}}
  \caption{SMC$^2$: $N=100$, $M=100$, $\alpha=0.99$, $s=0.1$.}
\end{subfigure}%
\hspace{0.5cm}
\begin{subfigure}{.3\textwidth}
  \centering
  \includegraphics[width=1\linewidth]{{{mean2_v_sims_5e-04_0.1_100_1000_0.9_0.5_}}}
  \caption{SMC$^2$: $N=100$, $M=1000$, $\alpha=0.9$, $s=0.1$.}
\end{subfigure}%
\hspace{0.5cm}
\begin{subfigure}{.3\textwidth}
  \centering
  \includegraphics[width=1\linewidth]{{{mean2_v_sims_5e-04_0.1_200_100_0.5_0.15_}}}
  \caption{ABC-SMC: $N=200$, $U=100$.}
\end{subfigure}
\caption{The estimated posterior mean of $\theta_2$ plotted against the total number of time steps used in Euler-Maruyama. Ground truth from ABC-MCMC is marked with a horizontal line.}
\end{figure}

\begin{figure}[H]
\centering
\begin{subfigure}{.3\textwidth}
  \centering
  \includegraphics[width=1\linewidth]{{{sd2_v_sims_5e-04_0.1_500_100_0.2_0.15_}}}
  \caption{DA-ABC-SMC: $N=500$, $U=100$, $A=100$, $s=0.1$.}
\end{subfigure}%
\hspace{0.5cm}
\begin{subfigure}{.3\textwidth}
  \centering
  \includegraphics[width=1\linewidth]{{{sd2_v_sims_5e-04_0.5_500_100_0.2_0.15_}}}
  \caption{DA-ABC-SMC: $N=500$, $U=100$, $A=100$, $s=0.5$.}
\end{subfigure}%
\hspace{0.5cm}
\begin{subfigure}{.3\textwidth}
  \centering
  \includegraphics[width=1\linewidth]{{{sd2_v_sims_5e-04_0.1_1000_100_0.1_0.15_}}}
  \caption{DA-ABC-SMC: $N=1000$, $U=100$, $A=100$, $s=0.1$.}
\end{subfigure}
\begin{subfigure}{.3\textwidth}
  \centering
  \includegraphics[width=1\linewidth]{{{sd2_v_sims_5e-04_0.5_1000_100_0.1_0.15_}}}
  \caption{DA-ABC-SMC: $N=1000$, $U=100$, $A=100$, $s=0.5$.}
\end{subfigure}%
\hspace{0.5cm}
\begin{subfigure}{.3\textwidth}
  \centering
  \includegraphics[width=1\linewidth]{{{sd2_v_sims_5e-04_0.1_1000_100_0.2_0.15_}}}
  \caption{DA-ABC-SMC: $N=1000$, $U=200$, $A=100$, $s=0.1$.}
\end{subfigure}%
\hspace{0.5cm}
\begin{subfigure}{.3\textwidth}
  \centering
  \includegraphics[width=1\linewidth]{{{sd2_v_sims_5e-04_0.1_1000_100_0.05_0.15_}}}
  \caption{DA-ABC-SMC: $N=1000$, $U=50$, $A=100$, $s=0.1$.}
\end{subfigure}
\begin{subfigure}{.3\textwidth}
  \centering
  \includegraphics[width=1\linewidth]{{{sd2_v_sims_5e-04_0.1_5000_100_0.02_0.15_}}}
  \caption{DA-ABC-SMC: $N=5000$, $U=100$, $A=100$, $s=0.1$.}
\end{subfigure}%
\hspace{0.5cm}
\begin{subfigure}{.3\textwidth}
  \centering
  \includegraphics[width=1\linewidth]{{{sd2_v_sims_5e-04_0.1_10000_100_0.01_0.15_}}}
  \caption{DA-ABC-SMC: $N=10000$, $U=100$, $A=100$, $s=0.1$.}
\end{subfigure}%
\hspace{0.5cm}
\begin{subfigure}{.3\textwidth}
  \centering
  \includegraphics[width=1\linewidth]{{{sd2_v_sims_5e-04_0.1_200_100_0.5_0.15_}}}
  \caption{ABC-SMC: $N=200$, $U=100$.}
\end{subfigure}
\caption{The estimated posterior standard deviation of $\theta_2$ plotted against the total number of time steps used in Euler-Maruyama. Ground truth from ABC-MCMC is marked with a horizontal line.}
\end{figure}

\begin{figure}[H]
\centering
\begin{subfigure}{.3\textwidth}
  \centering
  \includegraphics[width=1\linewidth]{{{mean3_v_sims_5e-04_0.1_500_100_0.2_0.15_}}}
  \caption{DA-ABC-SMC: $N=500$, $U=100$, $A=100$, $s=0.1$.}
\end{subfigure}%
\hspace{0.5cm}
\begin{subfigure}{.3\textwidth}
  \centering
  \includegraphics[width=1\linewidth]{{{mean3_v_sims_5e-04_0.5_500_100_0.2_0.15_}}}
  \caption{DA-ABC-SMC: $N=500$, $U=100$, $A=100$, $s=0.5$.}
\end{subfigure}%
\hspace{0.5cm}
\begin{subfigure}{.3\textwidth}
  \centering
  \includegraphics[width=1\linewidth]{{{mean3_v_sims_5e-04_0.1_1000_100_0.1_0.15_}}}
  \caption{DA-ABC-SMC: $N=1000$, $U=100$, $A=100$, $s=0.1$.}
\end{subfigure}
\begin{subfigure}{.3\textwidth}
  \centering
  \includegraphics[width=1\linewidth]{{{mean3_v_sims_5e-04_0.5_1000_100_0.1_0.15_}}}
  \caption{DA-ABC-SMC: $N=1000$, $U=100$, $A=100$, $s=0.5$.}
\end{subfigure}%
\hspace{0.5cm}
\begin{subfigure}{.3\textwidth}
  \centering
  \includegraphics[width=1\linewidth]{{{mean3_v_sims_5e-04_0.1_1000_100_0.2_0.15_}}}
  \caption{DA-ABC-SMC: $N=1000$, $U=200$, $A=100$, $s=0.1$.}
\end{subfigure}%
\hspace{0.5cm}
\begin{subfigure}{.3\textwidth}
  \centering
  \includegraphics[width=1\linewidth]{{{mean3_v_sims_5e-04_0.1_1000_100_0.05_0.15_}}}
  \caption{DA-ABC-SMC: $N=1000$, $U=50$, $A=100$, $s=0.1$.}
\end{subfigure}
\begin{subfigure}{.3\textwidth}
  \centering
  \includegraphics[width=1\linewidth]{{{mean3_v_sims_5e-04_0.1_5000_100_0.02_0.15_}}}
  \caption{DA-ABC-SMC: $N=5000$, $U=100$, $A=100$, $s=0.1$.}
\end{subfigure}%
\hspace{0.5cm}
\begin{subfigure}{.3\textwidth}
  \centering
  \includegraphics[width=1\linewidth]{{{mean3_v_sims_5e-04_0.1_10000_100_0.01_0.15_}}}
  \caption{DA-ABC-SMC: $N=10000$, $U=100$, $A=100$, $s=0.1$.}
\end{subfigure}%
\hspace{0.5cm}
\begin{subfigure}{.3\textwidth}
  \centering
  \includegraphics[width=1\linewidth]{{{mean3_v_sims_5e-04_0.1_100_100_0.9_0.5_}}}
  \caption{SMC$^2$: $N=100$, $M=100$, $\alpha=0.9$, $s=0.1$.}
\end{subfigure}
\begin{subfigure}{.3\textwidth}
  \centering
  \includegraphics[width=1\linewidth]{{{mean3_v_sims_5e-04_0.1_100_100_0.99_0.5_}}}
  \caption{SMC$^2$: $N=100$, $M=100$, $\alpha=0.99$, $s=0.1$.}
\end{subfigure}%
\hspace{0.5cm}
\begin{subfigure}{.3\textwidth}
  \centering
  \includegraphics[width=1\linewidth]{{{mean3_v_sims_5e-04_0.1_100_1000_0.9_0.5_}}}
  \caption{SMC$^2$: $N=100$, $M=1000$, $\alpha=0.9$, $s=0.1$.}
\end{subfigure}%
\hspace{0.5cm}
\begin{subfigure}{.3\textwidth}
  \centering
  \includegraphics[width=1\linewidth]{{{mean3_v_sims_5e-04_0.1_200_100_0.5_0.15_}}}
  \caption{ABC-SMC: $N=200$, $U=100$.}
\end{subfigure}
\caption{The estimated posterior mean of $\theta_3$ plotted against the total number of time steps used in Euler-Maruyama. Ground truth from ABC-MCMC is marked with a horizontal line.}
\end{figure}

\begin{figure}[H]
\centering
\begin{subfigure}{.3\textwidth}
  \centering
  \includegraphics[width=1\linewidth]{{{sd3_v_sims_5e-04_0.1_500_100_0.2_0.15_}}}
  \caption{DA-ABC-SMC: $N=500$, $U=100$, $A=100$, $s=0.1$.}
\end{subfigure}%
\hspace{0.5cm}
\begin{subfigure}{.3\textwidth}
  \centering
  \includegraphics[width=1\linewidth]{{{sd3_v_sims_5e-04_0.5_500_100_0.2_0.15_}}}
  \caption{DA-ABC-SMC: $N=500$, $U=100$, $A=100$, $s=0.5$.}
\end{subfigure}%
\hspace{0.5cm}
\begin{subfigure}{.3\textwidth}
  \centering
  \includegraphics[width=1\linewidth]{{{sd3_v_sims_5e-04_0.1_1000_100_0.1_0.15_}}}
  \caption{DA-ABC-SMC: $N=1000$, $U=100$, $A=100$, $s=0.1$.}
\end{subfigure}
\begin{subfigure}{.3\textwidth}
  \centering
  \includegraphics[width=1\linewidth]{{{sd3_v_sims_5e-04_0.5_1000_100_0.1_0.15_}}}
  \caption{DA-ABC-SMC: $N=1000$, $U=100$, $A=100$, $s=0.5$.}
\end{subfigure}%
\hspace{0.5cm}
\begin{subfigure}{.3\textwidth}
  \centering
  \includegraphics[width=1\linewidth]{{{sd3_v_sims_5e-04_0.1_1000_100_0.2_0.15_}}}
  \caption{DA-ABC-SMC: $N=1000$, $U=200$, $A=100$, $s=0.1$.}
\end{subfigure}%
\hspace{0.5cm}
\begin{subfigure}{.3\textwidth}
  \centering
  \includegraphics[width=1\linewidth]{{{sd3_v_sims_5e-04_0.1_1000_100_0.05_0.15_}}}
  \caption{DA-ABC-SMC: $N=1000$, $U=50$, $A=100$, $s=0.1$.}
\end{subfigure}
\begin{subfigure}{.3\textwidth}
  \centering
  \includegraphics[width=1\linewidth]{{{sd3_v_sims_5e-04_0.1_5000_100_0.02_0.15_}}}
  \caption{DA-ABC-SMC: $N=5000$, $U=100$, $A=100$, $s=0.1$.}
\end{subfigure}%
\hspace{0.5cm}
\begin{subfigure}{.3\textwidth}
  \centering
  \includegraphics[width=1\linewidth]{{{sd3_v_sims_5e-04_0.1_10000_100_0.01_0.15_}}}
  \caption{DA-ABC-SMC: $N=10000$, $U=100$, $A=100$, $s=0.1$.}
\end{subfigure}%
\hspace{0.5cm}
\begin{subfigure}{.3\textwidth}
  \centering
  \includegraphics[width=1\linewidth]{{{sd3_v_sims_5e-04_0.1_200_100_0.5_0.15_}}}
  \caption{ABC-SMC: $N=200$, $U=100$.}
\end{subfigure}
\caption{The estimated posterior standard deviation of $\theta_3$ plotted against the total number of time steps used in Euler-Maruyama. Ground truth from ABC-MCMC is marked with a horizontal line.}
\end{figure}

\section{Latent exponential random graph model\label{subsec:Latent-exponential-random}}

\begin{figure}
\centering
\begin{subfigure}{.5\textwidth}
  \centering
  \includegraphics[width=0.7\linewidth]{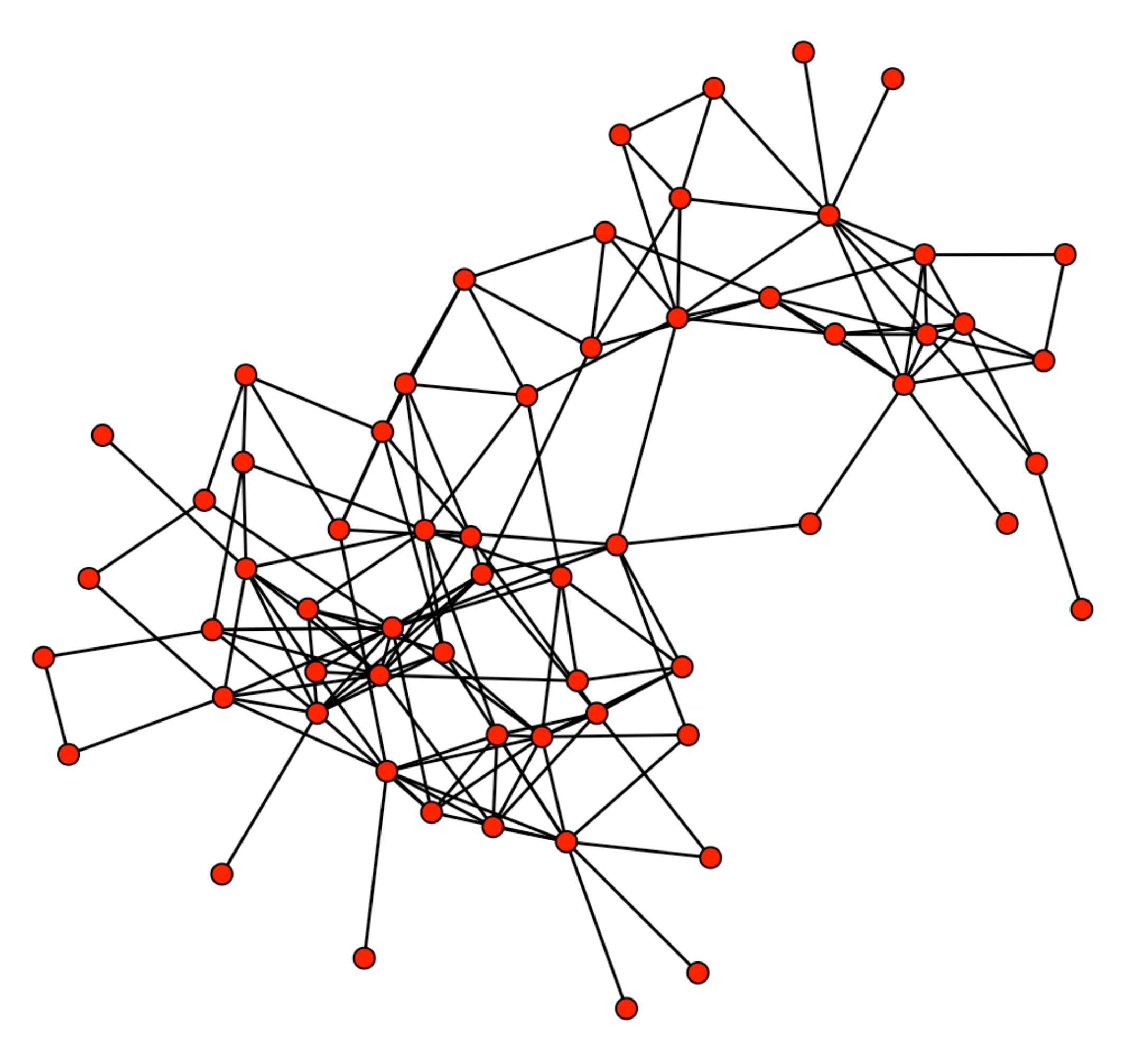}
  \caption{The Dolphin network.}
  \label{fig:The-Dolphin-network}
\end{subfigure}%
\begin{subfigure}{.5\textwidth}
  \centering
  \includegraphics[width=1\linewidth]{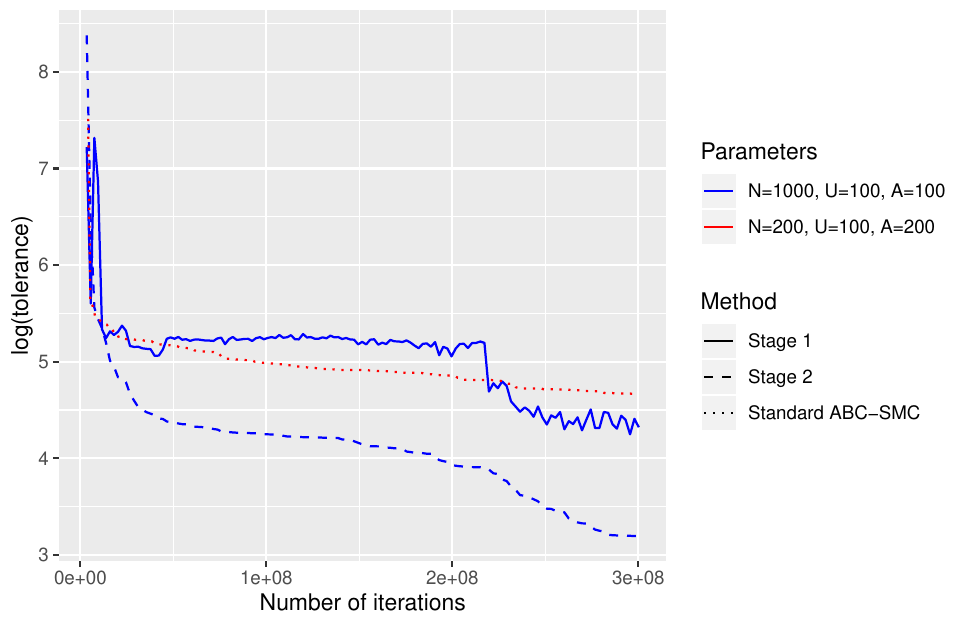}
  \caption{ABC tolerance plotted against the number of iterations of the TNT
sampler.}
  \label{fig:-plotted-against-2-1}
\end{subfigure}
\caption{DA-ABC-SMC applied to the latent ERGM.}
\end{figure}

An exponential random graph model (ERGM) is a~model for network data in which the global network structure
is modelled as having arisen through local interactions. In this section
we consider the situation in which the network is not directly observed,
thus $x^{h}$ is a~hidden network made up of a~random variable for
each edge which takes value 1 if the edge is present and 0 if it is
absent, and $y$ is a~noisy observation of this network. The ERGM
on $x^{h}$ is
\[
l\left(x^{h}\mid\theta_{x}\right)\propto\exp\left(\theta_{x}^{T}S\left(x^{h}\right)\right),
\]
with an intractable normalising constant, and our noisy observations are modelled by
\[
g\left(y_{i}\mid x_{i}^{h},\theta_{y}\right)\propto\exp\left(\theta_{y}\left(2x_{i}^{h}-1\right)\left(2y_{i}-1\right)\right)
\]
where the normalising constant is tractable. We studied the Dolphin
network (figure \ref{fig:The-Dolphin-network}) \citep{Lusseau2003},
as also analysed in \citet{Caimo2011} where the network is treated
as directly observed, and used the same summary statistics and priors
as in this paper. The  \texttt{igraph} package \citep{igraph} was used to load this data into R. We used the statistics
\begin{eqnarray*}
S_{1}(x^h)=\sum_{i<j}x^h_{ij} & \text{the number of edges}\\
S_{2}(x^h)=\exp\left(\phi_{u}\right)\sum_{i=1}^{n-1}\left\{ 1-\left(1-\exp\left(-\phi_{u}\right)\right)^{i}\right\} D_{i}\left(x^h\right) & \text{geometrically weighted degree}\\
S_{3}(x^h)=\exp\left(\phi_{v}\right)\sum_{i=1}^{n-2}\left\{ 1-\left(1-\exp\left(-\phi_{v}\right)\right)^{i}\right\} EP_{i}\left(x^h\right) & \text{geometrically weighted edgewise shared partner}
\end{eqnarray*}
\foreignlanguage{british}{with $\phi_{u}=\phi_{v}=0.8$, the prior
on $\theta_{x}=(\theta_{1},\theta_{2},\theta_{3})$ and $\theta_y$ was $\left(\theta_{1},\theta_{2},\theta_{3},\theta_{y}\right)\sim\mathcal{N}(0,30I_{4})$;
and we used the Euclidean distance to compare simulated with observed
statistics. The \texttt{ergm} package \citep{ergm} in R was
used to simulate from $l\left(\cdot\mid\theta_{x}\right)$, which
uses the \textquotedblleft tie no tie\textquotedblright{} (TNT) sampler
and the expensive simulator used $15,000$ iterations. }Our DA-ABC-SMC
algorithm used $U=A=100$, and again the MCMC proposal was taken to
be a~Gaussian distribution centred at the current particle, with covariance
given by the sample covariance of the particles from the previous
iteration.

We ran DA-ABC-SMC for $N=1,000$, and a~cheap simulator having $B=1,500$
(after exploratory runs suggested that this would be enough iterations
to provide a~useful DA proposal) and compared the results with standard
ABC-SMC with the same configuration as in the previous section (both
using $3\times10^{8}$ iterations of the TNT sampler). Figure \ref{fig:-plotted-against-2-1}
shows the results from the two algorithms, this time showing the sequence
$\epsilon_{1,t}$ alongside the sequence $\epsilon_{2,t}$. Again
we observe that the tolerance in DA-ABC-SMC reduces faster than standard ABC-SMC, and that the
tolerance $\epsilon_{1,t}$ changes adaptively. This data has not previously been studied using
a latent ERGM. Using particle MCMC as in \citet{Everitt2012} would
require at every MCMC iteration to run an SMC sampler to integrate
out the latent ERGM space, which consists of 1891 binary edge variables.
We might expect that many SMC particles would be required to produce
low variance marginal likelihood estimates, leading to a~high computational
cost. However, the acceptance rate was very low towards the end of
our ABC runs, suggesting that a~very large computational cost would
be required to reduce $\epsilon_{2,t}$ to be close to zero.



\bibliographystyle{mychicago}
\bibliography{da_paper}

\end{document}